\documentclass[prd,aps,twocolumn,preprintnumbers, showpacs,
nofootinbib,superscriptaddress,notitlepage,floatfix]{revtex4-1}
\usepackage{amsthm,amsmath,amssymb}
\usepackage{mathrsfs}
\usepackage{graphicx}
\usepackage{slashed}
\usepackage{multirow}
\usepackage[colorlinks, citecolor=blue,anchorcolor=blue,menucolor=blue,
linkcolor=blue,filecolor=blue,runcolor=blue,urlcolor=blue,frenchlinks=true]{hyperref}
\usepackage[normalem]{ulem}
\usepackage{xcolor}

\begin{document}

\title{Gluon Parton Distribution of the Nucleon from 2+1+1-Flavor Lattice QCD in the Physical-Continuum Limit}

\author{Zhouyou Fan}
\affiliation{Department of Physics and Astronomy, Michigan State University, East Lansing, MI 48824}

\author{William Good}
\affiliation{Department of Physics and Astronomy, Michigan State University, East Lansing, MI 48824}
\affiliation{Department of Computational Mathematics,
  Science and Engineering, Michigan State University, East Lansing, MI 48824}

\author{Huey-Wen Lin}
\affiliation{Department of Physics and Astronomy, Michigan State University, East Lansing, MI 48824}
\affiliation{Department of Computational Mathematics,
  Science and Engineering, Michigan State University, East Lansing, MI 48824}

\preprint{MSUHEP-22-033}

\begin{abstract}
We present the first physical-continuum limit $x$-dependent nucleon gluon distribution from lattice QCD using the pseudo-PDF approach, on lattice ensembles with $2+1+1$ flavors of highly improved staggered quarks (HISQ), generated by MILC Collaboration.
We use clover fermions for the valence action on three lattice spacings $a \approx 0.9$, 0.12 and 0.15~fm and three pion masses $M_\pi \approx 220$, 310 and 690~MeV, with nucleon two-point measurements numbering up to $O(10^6)$ and nucleon boost momenta up to 3~GeV. We study the lattice-spacing and pion-mass dependence of the reduced pseudo-ITD matrix elements obtained from the lattice calculation, then extrapolate them to the continuum-physical limit before extracting  $xg(x)/\langle x \rangle_g$.
We use the gluon momentum fraction $\langle x \rangle_g$ calculated from the same ensembles to determine the nucleon gluon unpolarized PDF $xg(x)$ for the first time entirely through lattice-QCD simulation. We compare our results with previous single-ensemble lattice calculations, as well as selected global fits.
\end{abstract}

\maketitle

\section{Introduction}
Many precision phenomenology and theoretical predictions for hadron colliders rely on accurate estimates of the uncertainty in Standard-Model (SM) predictions.
Among these predictions, the parton distribution functions (PDFs), the nonperturbative functions quantifying probabilities for finding quarks and gluons in hadrons with particular momentum fraction, are particularly important inputs in high-energy scattering~\cite{Harland-Lang:2014zoa,Dulat:2015mca,Abramowicz:2015mha,Accardi:2016qay,Alekhin:2017kpj,Ball:2017nwa,Hou:2019efy,Bailey:2019yze,Bailey:2020ooq,Ball:2021leu,ATLAS:2021vod}.
The gluon PDF $g(x)$ needs to be known precisely to calculate the cross section for these processes in $pp$ collisions, such as the cross section for Higgs-boson production and jet production at the Large Hadron Collider (LHC)~\cite{CMS:2012nga,Kogler:2018hem}, and direct $J/\psi$ photoproduction at Jefferson Lab~\cite{mammeiproposal}.
The future U.S.-based Electron-Ion Collider (EIC)~\cite{Accardi:2012qut}, planned to be built at Brookhaven National Lab, will further our knowledge of the gluon PDF~\cite{Arrington:2021biu,Aguilar:2019teb,AbdulKhalek:2021gbh}.
In Asia, the Electron-Ion Collider in China (EicC)~\cite{Anderle:2021wcy} is also planned to impact the gluon and sea-quark distributions.
Although significant efforts to extract the gluon distribution $g(x)$ have been made in the last decade, there are still problems in obtaining a precise $g(x)$ in the large-$x$. 

Lattice quantum chromodynamics (QCD) is a nonperturbative theoretical method for calculating QCD quantities that has full systematic control.
Calculations of $x$-dependent hadron structure in lattice QCD have multiplied since the proposal of Large-Momentum Effective Theory (LaMET)~\cite{Ji:2013dva,Ji:2014gla,Ji:2017rah}.
Many lattice works have been done on nucleon and meson PDFs, and generalized parton distributions (GPDs) based on the quasi-PDF approach~\cite{Lin:2013yra,Lin:2014zya,Chen:2016utp,Lin:2017ani,Alexandrou:2015rja,Alexandrou:2016jqi,Alexandrou:2017huk,Chen:2017mzz,Alexandrou:2018pbm,Chen:2018xof,Chen:2018fwa,Alexandrou:2018eet,Lin:2018qky,Fan:2018dxu,Liu:2018hxv,Wang:2019tgg,Lin:2019ocg,Chen:2019lcm,Lin:2020reh,Chai:2020nxw,Bhattacharya:2020cen,Lin:2020ssv,Zhang:2020dkn,Li:2020xml,Fan:2020nzz,Gao:2020ito,Lin:2020fsj,Zhang:2020rsx,Alexandrou:2020qtt,Alexandrou:2020zbe,Lin:2020rxa,Gao:2021hxl,Lin:2020rut}.
Alternative approaches to lightcone PDFs in lattice QCD are
the Compton-amplitude approach (or ``OPE without OPE'')~\cite{Aglietti:1998ur,Martinelli:1998hz,Dawson:1997ic,Capitani:1998fe,Capitani:1999fm,Ji:2001wha,Detmold:2005gg,Braun:2007wv,Chambers:2017dov,Detmold:2018kwu,QCDSF-UKQCD-CSSM:2020tbz,Horsley:2020ltc,Detmold:2021uru}, 
the ``hadronic-tensor approach''~\cite{Liu:1993cv,Liu:1998um,Liu:1999ak,Liu:2016djw,Liu:2017lpe,Liu:2020okp},
the ``current-current correlator''~\cite{Braun:2007wv,Ma:2017pxb,Bali:2017gfr,Bali:2018spj,Joo:2020spy,Gao:2020ito,Sufian:2019bol,Sufian:2020vzb}
and the pseudo-PDF approach~\cite{Radyushkin:2017cyf,Orginos:2017kos,Karpie:2017bzm,Karpie:2018zaz,Karpie:2019eiq,Joo:2019jct,Joo:2019bzr,Radyushkin:2018cvn,Zhang:2018ggy,Izubuchi:2018srq,Joo:2020spy,Bhat:2020ktg,Fan:2020cpa,Sufian:2020wcv,Karthik:2021qwz,HadStruc:2021wmh,Fan:2021bcr,HadStruc:2022yaw}. 
A few works have started to include lattice-QCD systematics, such as finite-volume effects, in their calculations~\cite{Lin:2019ocg,Sufian:2020vzb}.
However, most these calculations are still, at the current stage, done with a single lattice spacing.
Most lattice calculations of PDFs use next-to-leading-order (NLO) matching or, equivalently, NLO Wilson coefficients~\cite{Xiong:2013bka,Ma:2014jla,Ji:2017rah, Ji:2020ect}, and some lattice calculations of the valence pion PDF~\cite{Gao:2021dbh} have incorporated NNLO matching~\cite{Chen:2020ody,Li:2020xml}.
More work is needed to reduce high-twist systematics and improve the lattice determination of small-$x$ and antiquark PDFs with very large boost momenta.

Recently, progress has been made in the most-calculated isovector quark distribution of nucleon by MSULat~\cite{Lin:2020fsj}, ETMC~\cite{Alexandrou:2020qtt} and HadStruc Collaborations~\cite{Karpie:2021pap}, who studied lattice-spacing dependence.
MSULat studied three lattice spacings (0.09, 0.12 and 0.15~fm) and pion masses (135, 220, 310~MeV) and performed a simultaneous continuum-physical extrapolation using a third-order $z$-expansion on renormalized LaMET matrix elements~\cite{Lin:2020fsj} with nucleon boost momenta around 2.2 and 2.6~GeV. 
ETMC also uses three lattice spacings, 0.06, 0.08, and 0.09~fm, but with heavier pion mass (370~MeV) and investigated the continuum extrapolation of the data on renormalized LaMET matrix elements with boost momentum around 1.8~GeV~\cite{Alexandrou:2020qtt}.
HadStruc Collaboration studied three lattice spacings, 0.048, 0.065, and 0.075~fm with two-flavor 440-MeV lattice ensembles using the continuum pseudo--Ioffe-time distribution (ITD)~\cite{Karpie:2021pap}. 
Most of the works above found mild nonzero dependence on lattice spacing (varying with the Wilson-link displacement) in the nucleon case for LaMET or pseudo-ITD matrix elements. 

In contrast with the quark PDFs, the gluon PDFs calculations are less calculated, due to their notoriously noisier matrix elements on the lattice.
To date, there have only been a few exploratory gluon-PDF calculations for unpolarized nucleon~\cite{Fan:2018dxu,Fan:2020cpa,HadStruc:2021wmh}, pion~\cite{Fan:2021bcr} and kaon~\cite{Salas-Chavira:2021wui}, and polarized nucleon~\cite{HadStruc:2022yaw} using the pseudo-PDF~\cite{Balitsky:2019krf}  and quasi-PDF~\cite{Zhang:2018diq,Wang:2019tgg} methods.
Most of these calculations, like many exploratory lattice calculations, are done only using one lattice spacing at heavy pion mass.

In this work, we report the first continuum-limit unpolarized gluon PDF of nucleon study using three lattice spacings: 0.09, 0.12 and 0.15~fm with pion mass ranging from 220 to 700~MeV using the pseudo-PDF method. 
The remainder of this paper is organized as follows: 
In Sec.~\ref{sec:cal-details}, we present the procedure for how the lattice correlators are calculated and analyzed to extract the ground-state matrix elements for the pseudo-PDF method.
We then study the lattice-spacing dependence of matrix elements at 310 and 700-MeV pion mass in Sec.~\ref{sec:results}, checking both $O(a)$ and $O(a^2)$ forms using multiple continuum-extrapolation strategies.
We perform a physical-continuum extrapolation to obtain continuum reduced pseudo-ITDs (RpITDs) matrix element, before final determination of the nucleon unpolarized gluon PDF $xg(x)$ is obtained from the $xg(x)/\langle x \rangle_g$ and $\langle x \rangle_g$ results.
Using the gluon momentum fraction calculated on the same ensemble, we obtain the gluon PDF and compare with the phenomenological global-fit PDF results.
We consider the quark-mixing systematics, but they are found to be small.
The final conclusion and future outlook can be found in Sec.~\ref{sec:summary}.

\section{Lattice Setup, Correlators and Matrix Elements}
\label{sec:cal-details}

This calculation is carried out using four ensembles with $N_f = 2+1+1$ highly improved staggered quarks (HISQ)~\cite{Follana:2006rc}, generated by the MILC Collaboration~\cite{Bazavov:2012xda}, with three different lattice spacings ($a\approx 0.9$, 0.12 and 0.15~fm) and three pion masses (220, 310, 690~MeV);
see Table~\ref{table-data} for more details.
We apply five steps of hypercubic (HYP) smearing~\cite{Hasenfratz:2001hp} to the gauge links to reduce short-distance noise.
Wilson-clover fermions are used in the valence sector, and the valence-quark masses are tuned to reproduce the lightest light and strange sea pseudoscalar meson masses (which correspond to pion masses 310 and 690~MeV, respectively).
A similar setup is used by PNDME collaboration~\cite{Mondal:2020cmt,Park:2020axe,Jang:2019jkn,Jang:2019vkm,Gupta:2018lvp,Lin:2018obj,Gupta:2018qil,Gupta:2017dwj,Rajan:2017lxk,Bhattacharya:2015wna,Bhattacharya:2015esa,Bhattacharya:2013ehc} with local operators, such as isovector and flavor-diagonal charges, form factors and moments;
the results from this mixed-action setup are consistent with the same physical quantities calculated using different fermion actions~\cite{Kronfeld:2019nfb,Lin:2017snn,Lin:2020rut,Lin:2022nnj,FlavourLatticeAveragingGroup:2019iem,FlavourLatticeAveragingGroupFLAG:2021npn}.

\begin{table}[!htbp]
\centering
\begin{tabular}{|c|c|c|c|c|}
\hline
  Ensemble & a09m310 & a12m220 & a12m310 & a15m310 \\
\hline
  $a$ (fm) & $0.0888(8)$ & $0.1184(10)$ & $0.1207(11)$  & $0.1510(20)$ \\
\hline
  $L^3\times T$ & $32^3\times 96$ & $32^3\times 64$ & $24^3\times 64$ & $16^3\times 48$ \\
\hline
  $M_{\pi}^\text{val}$ (GeV) & $0.313(1)$ & $0.2266(3)$ & $0.309(1)$ & $0.319(3)$\\
\hline
  $M_{\eta_s}^\text{val}$ (GeV) & 0.698(7) & N/A & $0.6841(6)$ & $0.687(1)$\\
\hline
  $P_z$ (GeV)  & $[0,3.05]$ & $[0,2.29]$ &  $[0,2.14]$ &  $[0,2.56]$   \\
\hline
$N_\text{cfg}$ & 1009 & 957 & 1013 & 900   \\
\hline
  $N_\text{meas}^\text{2pt}$ & 387,456 &  1,466,944  & 324,160 & 259,200  \\
\hline 
$t_\text{sep}$ & $[6,10]$ &  $[6,10]$   &  $[5,9]$ & $[4,8]$  \\
\hline
\end{tabular}
\caption{ 
Lattice spacing $a$, valence pion mass ($M_\pi^\text{val}$) and $\eta_s$ mass ($M_{\eta_s}^\text{val}$), lattice size ($L^3\times T$), number of configurations ($N_\text{cfg}$), number of total two-point correlator measurements ($N_\text{meas}^\text{2pt}$), and  source-sink separation times $t_\text{sep}$ used in the three-point correlator fits of $N_f=2+1+1$ clover valence fermions on HISQ ensembles generated by the MILC Collaboration and analyzed in this study.
}
\label{table-data}
\end{table}

\begin{figure*}[htbp]
\centering
\includegraphics[width=0.3333\textwidth]{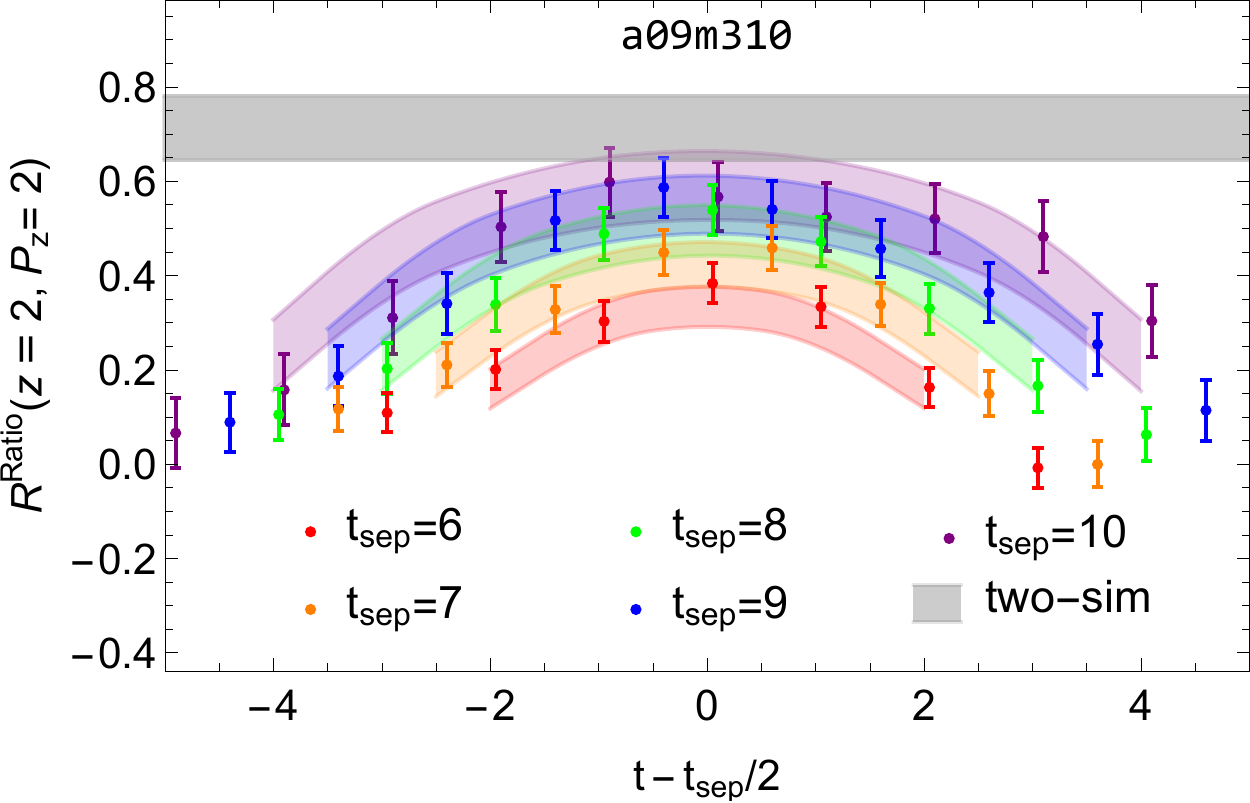}
\centering
\centering
\includegraphics[width=0.198\textwidth]{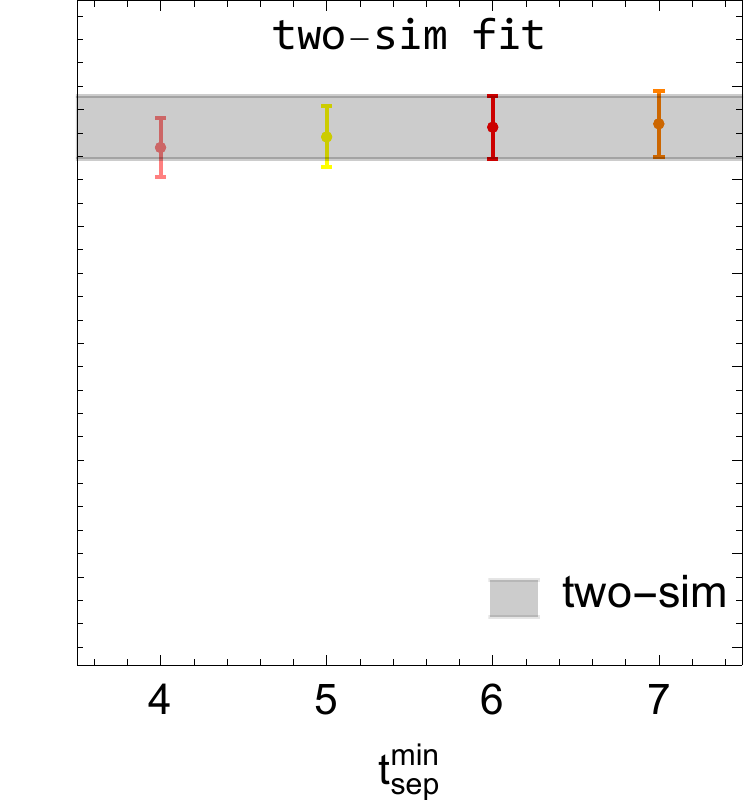}
\centering
\includegraphics[width=0.200\textwidth]{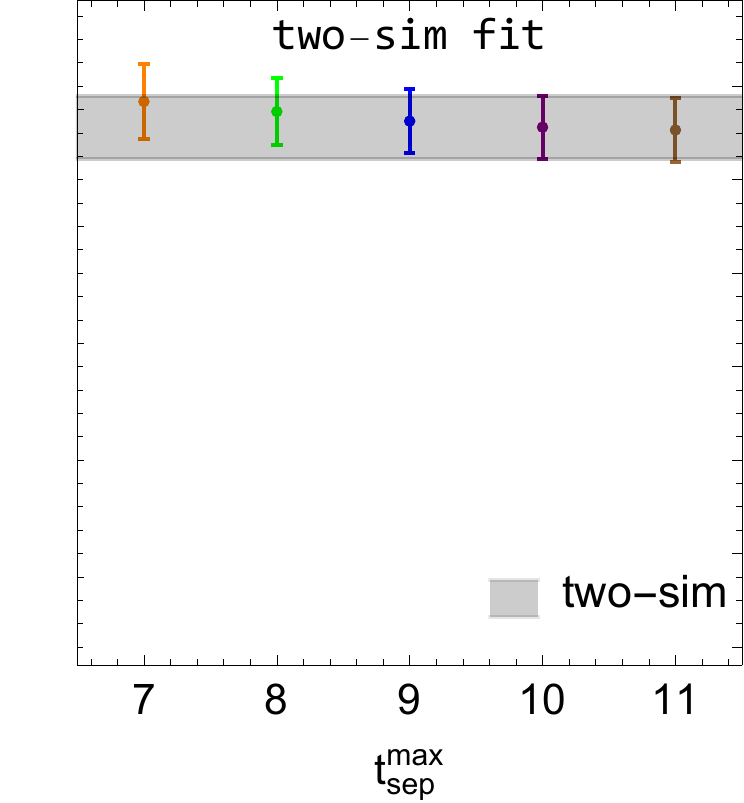}
\centering
\includegraphics[width=0.3333\textwidth]{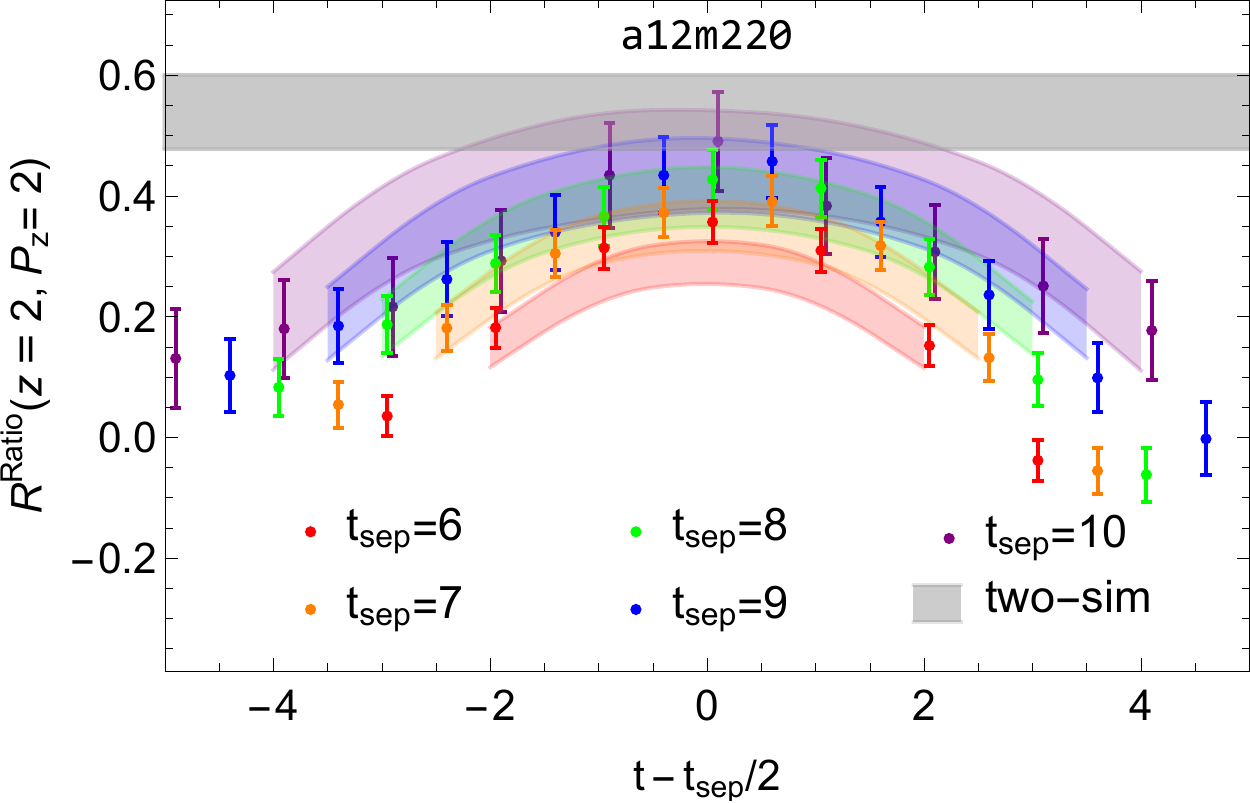}
\centering
\centering
\includegraphics[width=0.198\textwidth]{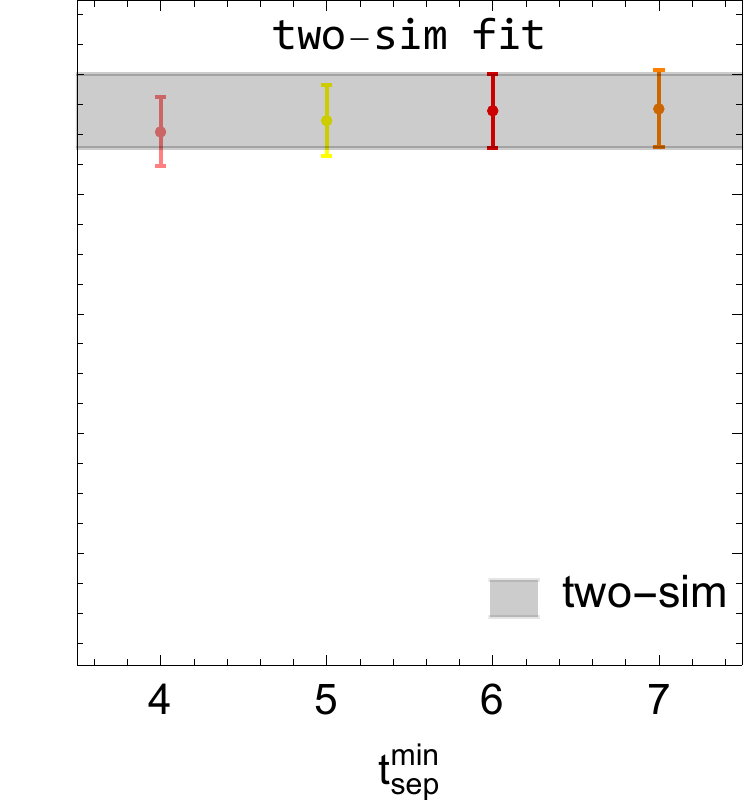}
\centering
\includegraphics[width=0.200\textwidth]{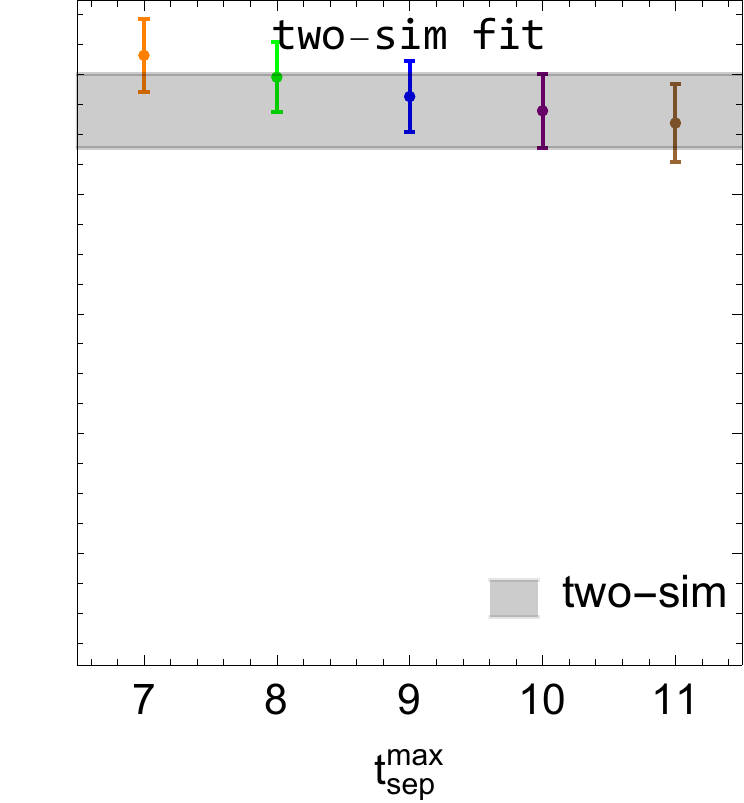}
\centering
\includegraphics[width=0.3333\textwidth]{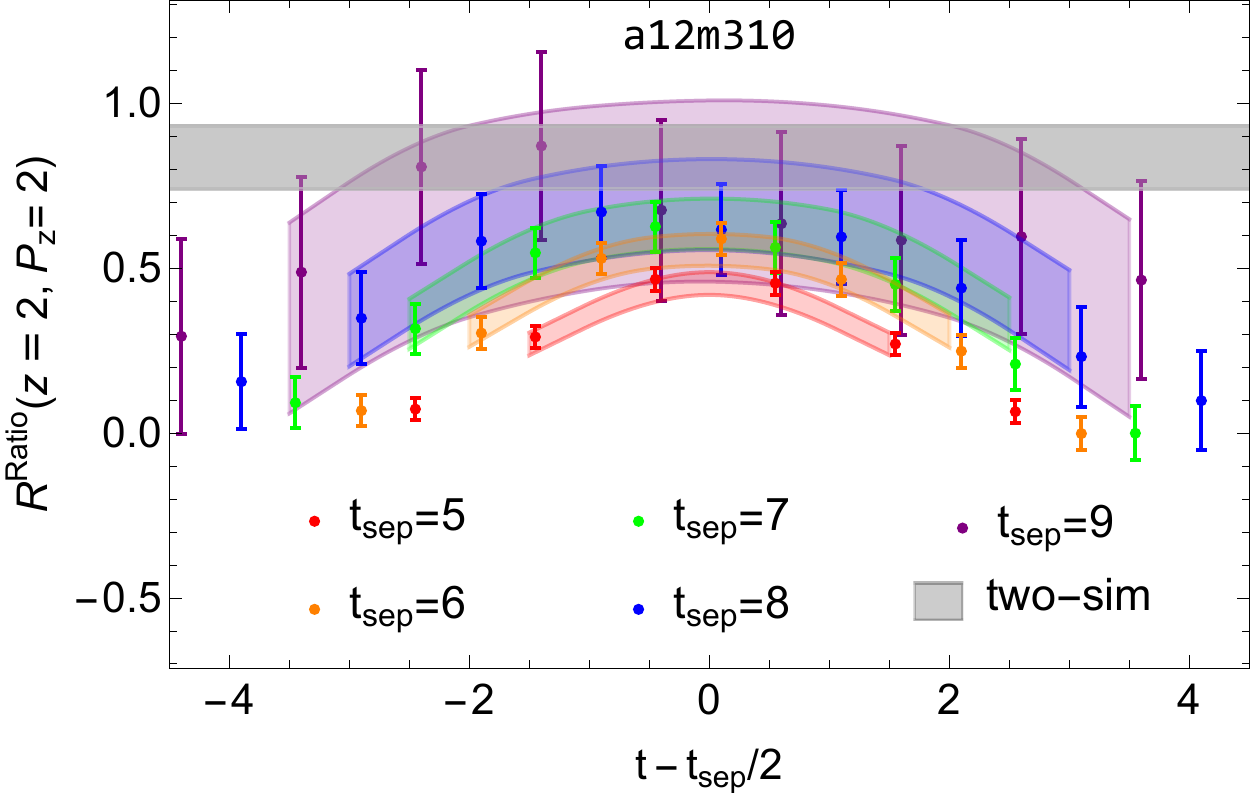}
\centering
\centering
\includegraphics[width=0.198\textwidth]{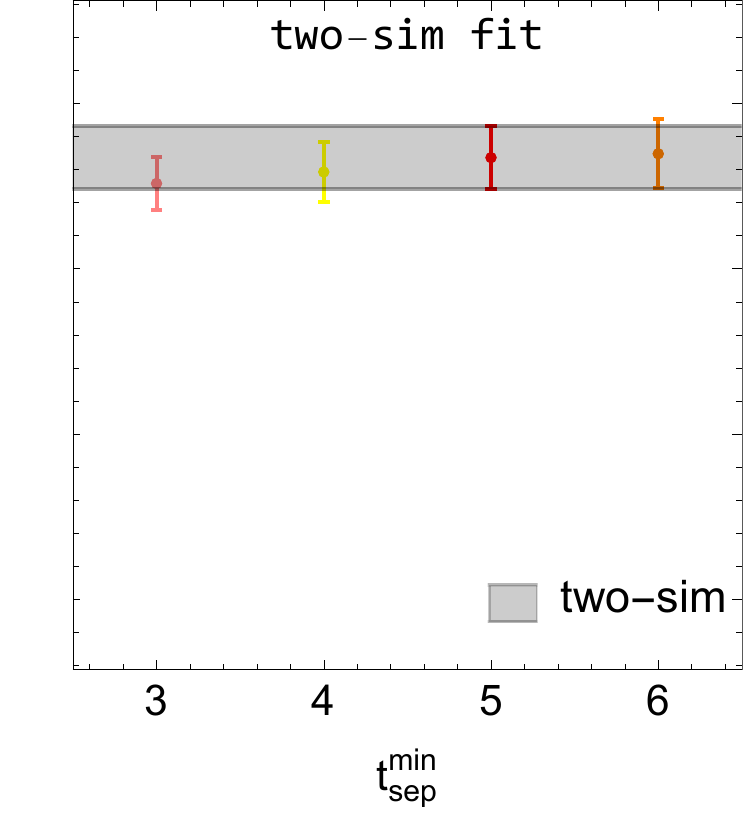}
\centering
\includegraphics[width=0.200\textwidth]{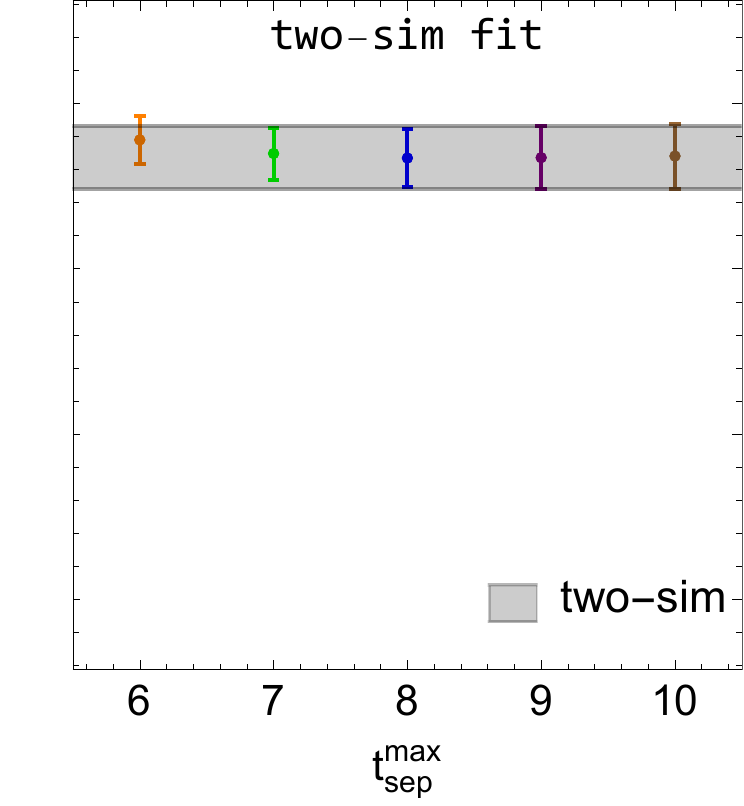}
\centering
\includegraphics[width=0.3333\textwidth]{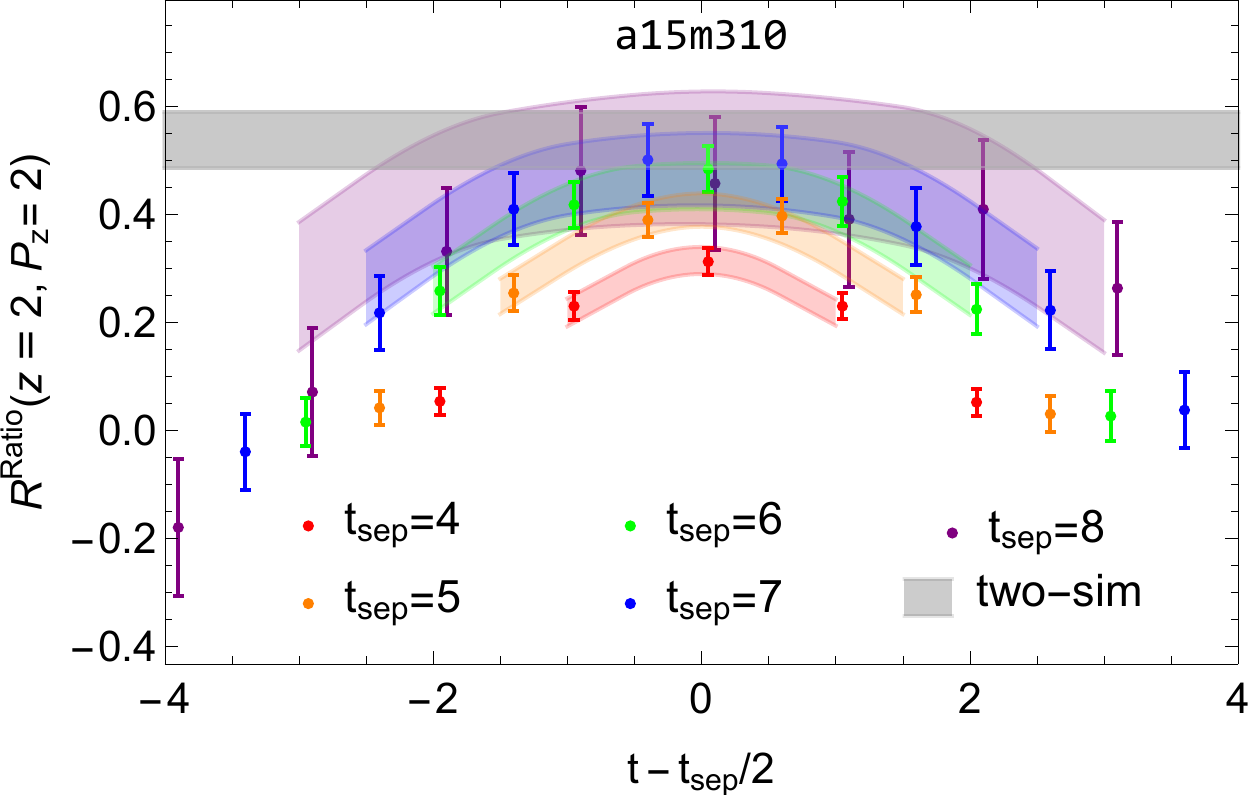}
\centering
\centering
\includegraphics[width=0.198\textwidth]{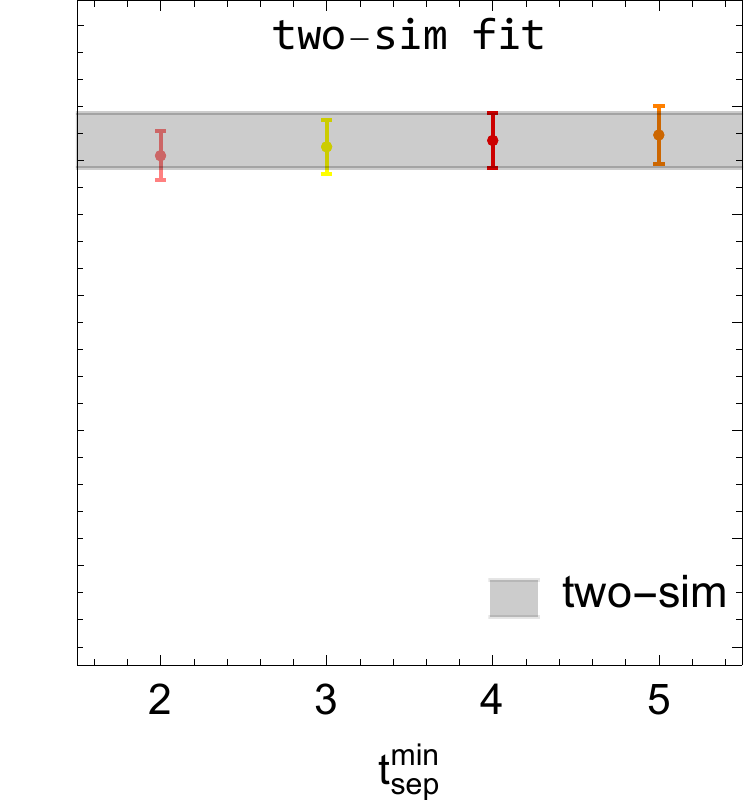}
\centering
\includegraphics[width=0.200\textwidth]{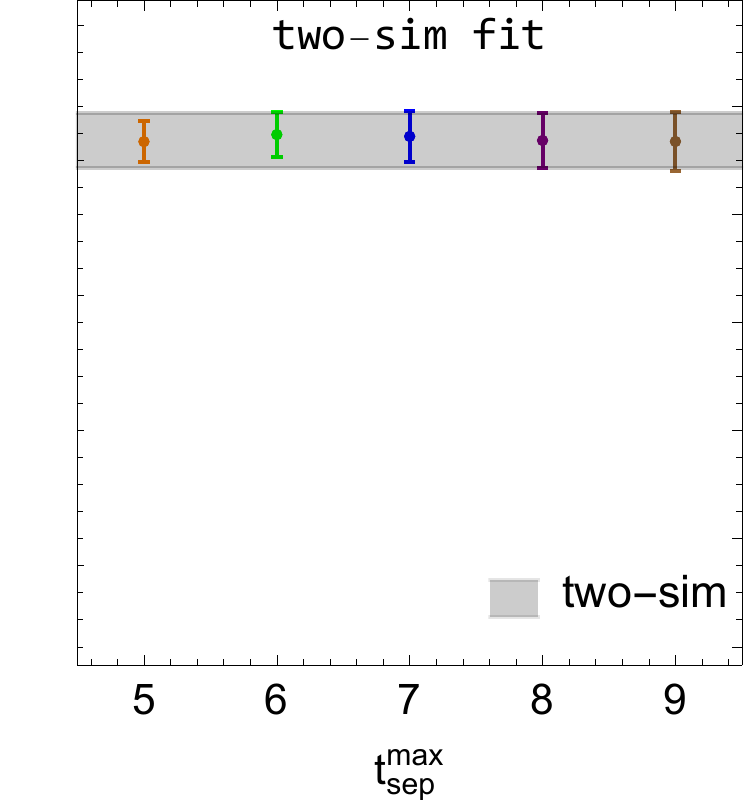}
\caption{ 
Example ratio plots (left) and two-sim fits (last 2 columns) of the light nucleon correlators at pion masses $M_\pi \approx \{310, 220,310, 310\}$~MeV from the a09m310, a12m220, a12m310 and a15m310 ensembles.
The gray band shown on each plot is the extracted ground-state matrix element from the two-sim fit that we use as our best value.
From left to right, the columns are:
the ratio of the three-point to two-point correlators with the reconstructed fit bands from the two-sim fit using the final $t_\text{sep}$ inputs, shown as functions of $t-t_\text{sep}/2$,
the one-state fit results for the three-point correlators at different $t_\text{sep}$ values,
the two-sim fit results using $t_\text{sep}\in[t_\text{sep}^\text{min},t_\text{sep}^\text{max}]$ varying $t_\text{sep}^\text{min}$ and $t_\text{sep}^\text{max}$. }
\label{fig:LRatio-fitcomp}
\end{figure*}

\begin{figure*}[htbp]
\centering
\includegraphics[width=0.3333\textwidth]{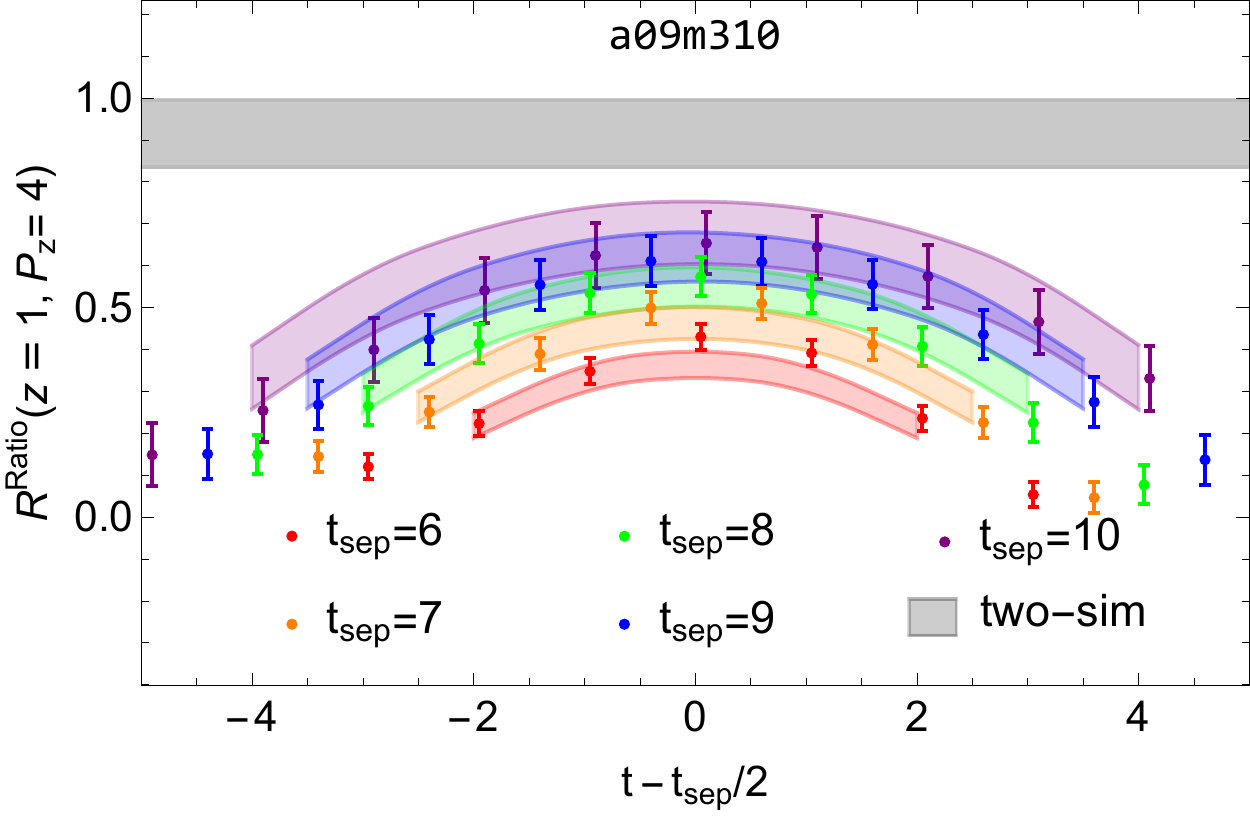}
\centering
\centering
\includegraphics[width=0.198\textwidth]{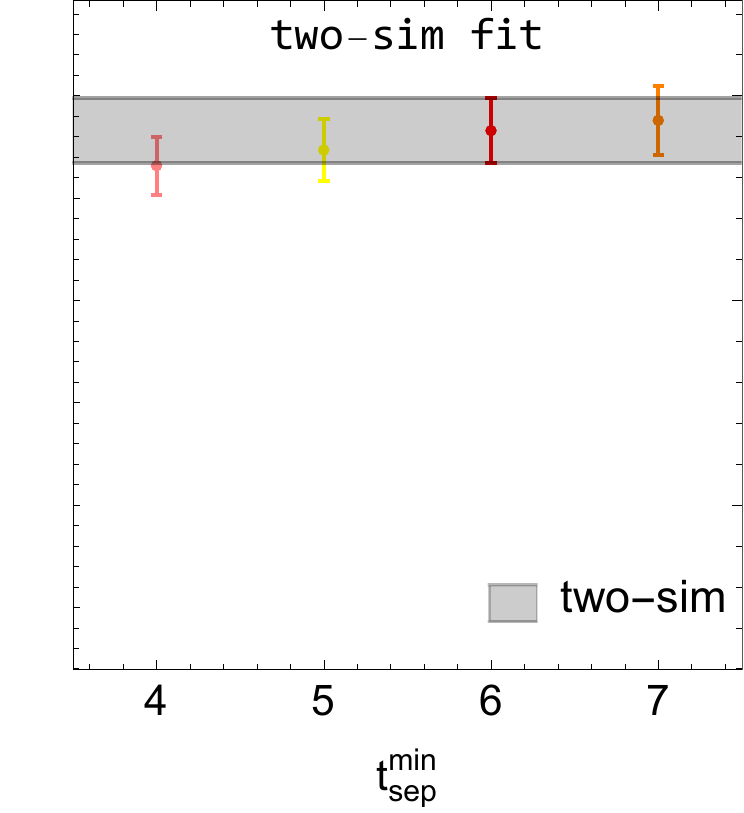}
\centering
\includegraphics[width=0.200\textwidth]{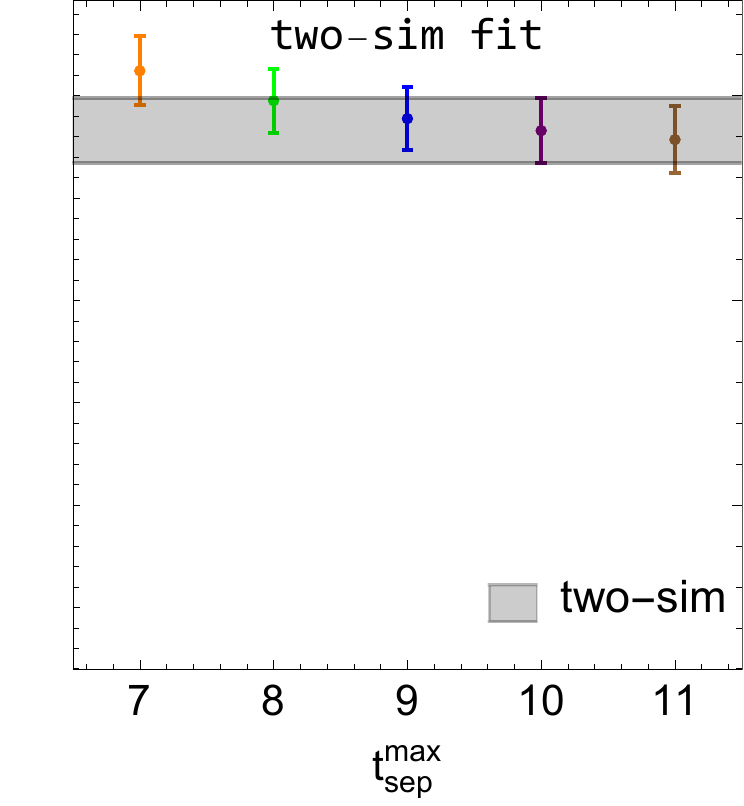}
\centering
\includegraphics[width=0.3333\textwidth]{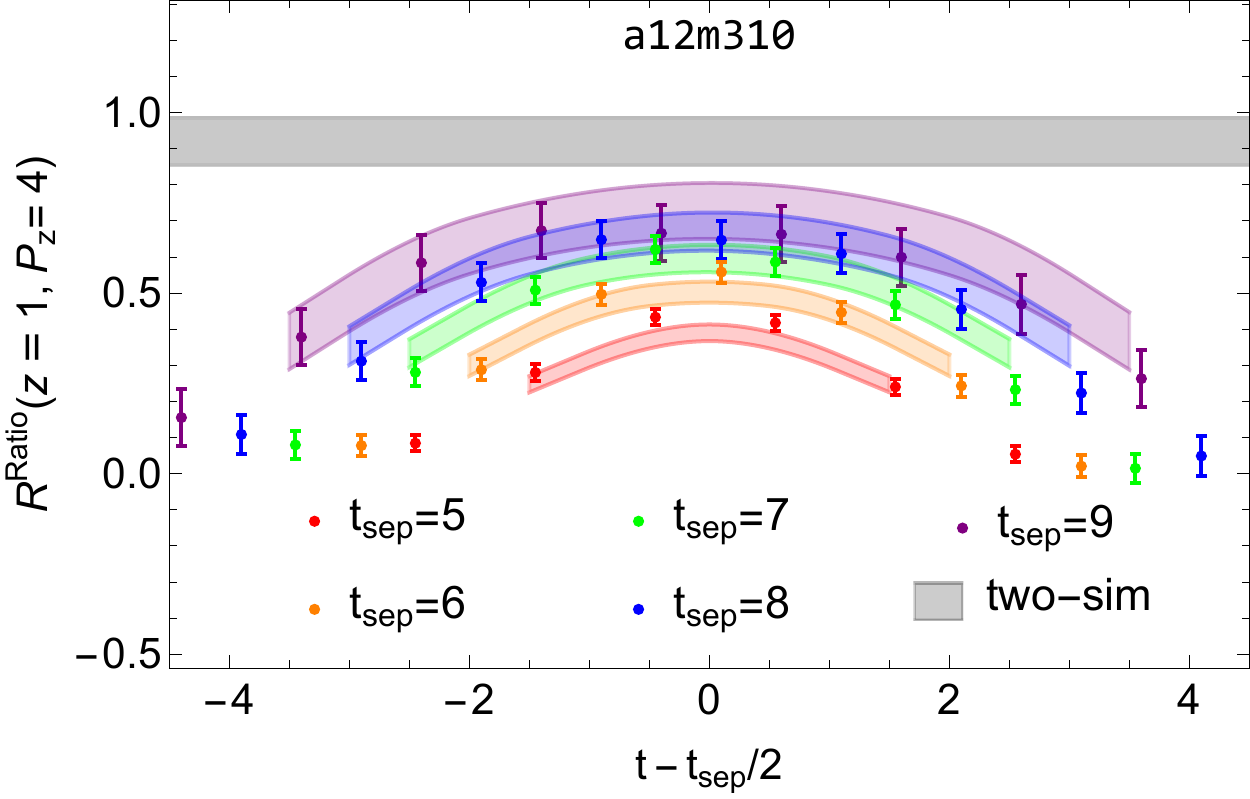}
\centering
\includegraphics[width=0.198\textwidth]{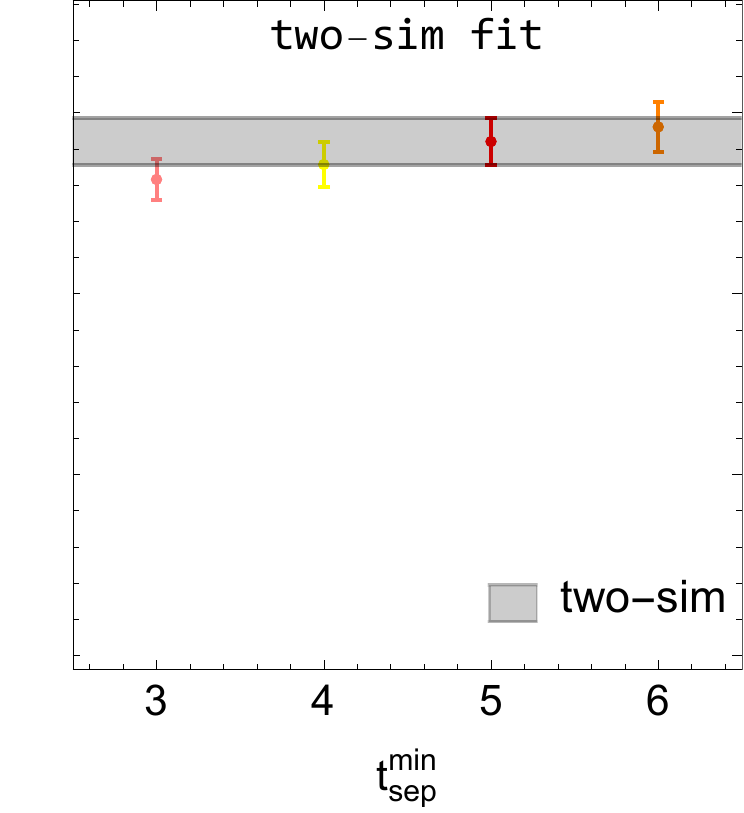}
\centering
\includegraphics[width=0.200\textwidth]{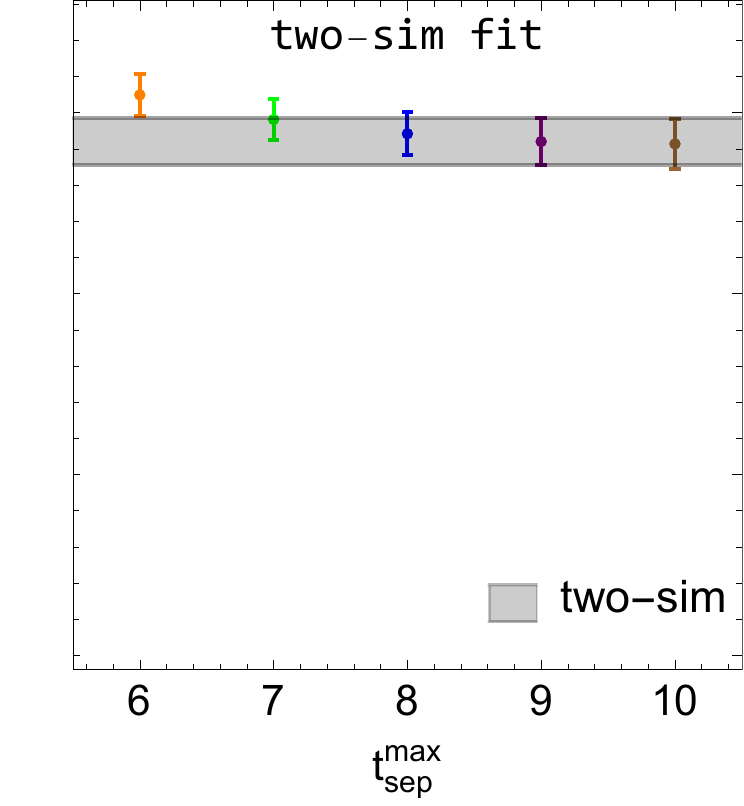}
\centering
\includegraphics[width=0.3333\textwidth]{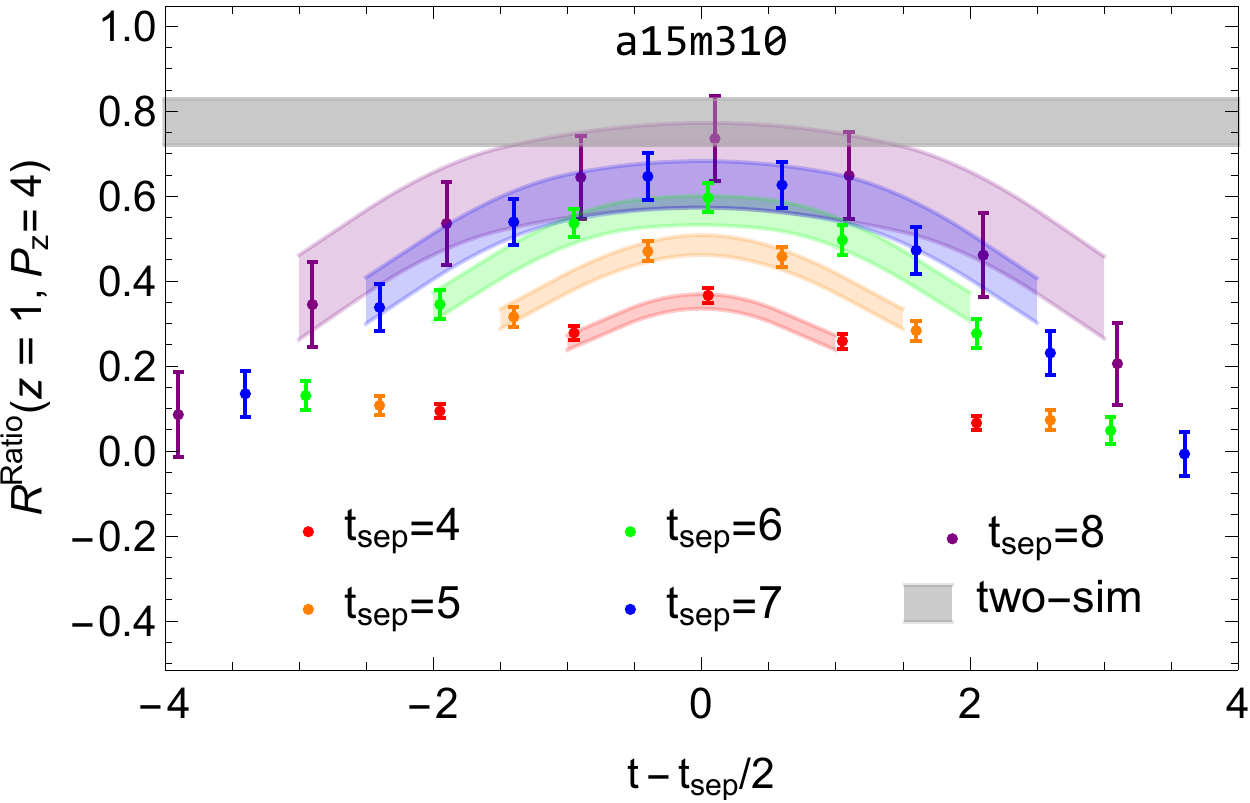}
\centering
\includegraphics[width=0.198\textwidth]{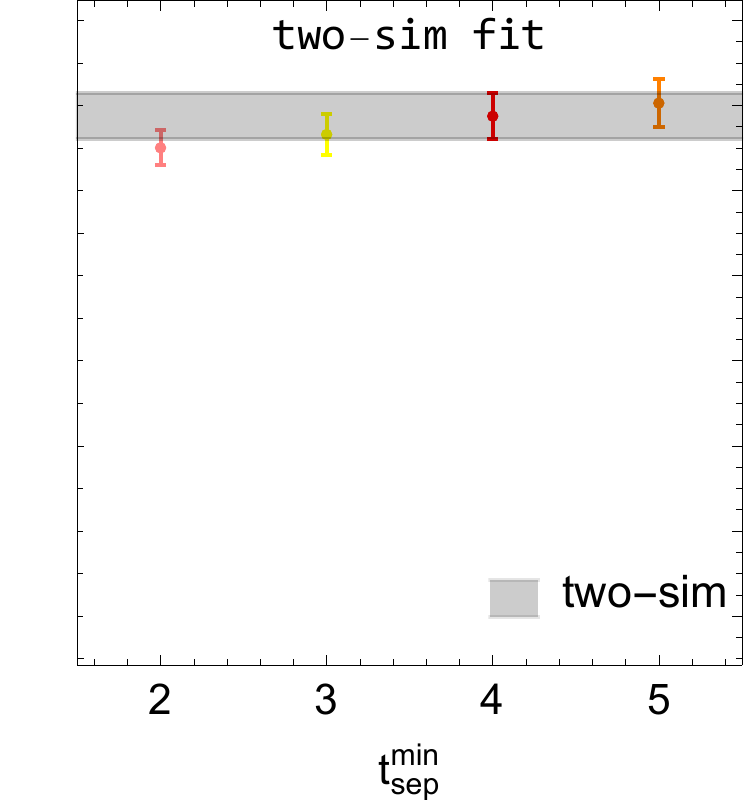}
\centering
\includegraphics[width=0.200\textwidth]{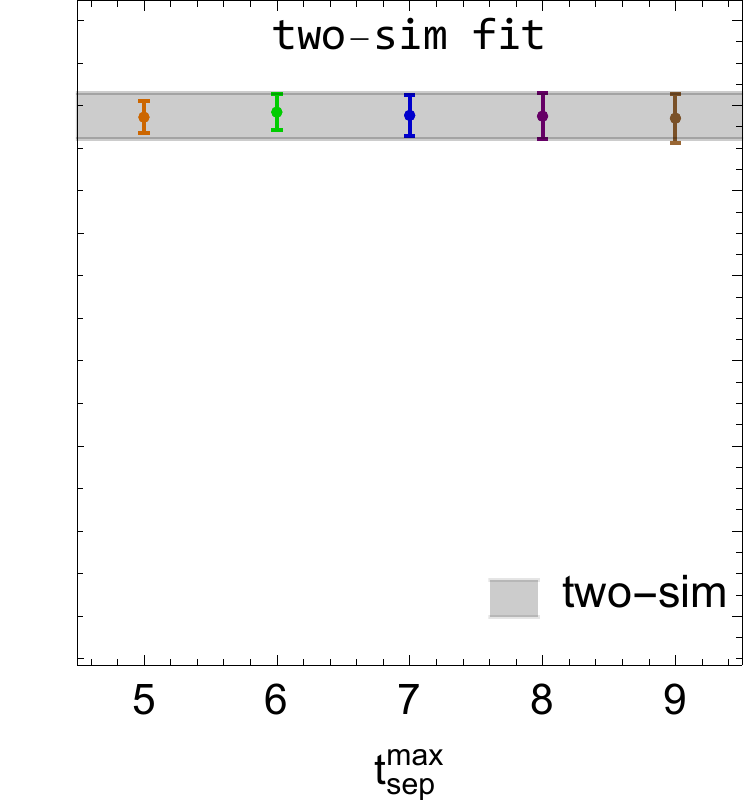}
\caption{ 
Example ratio plots (left),
and two-sim fits (last 2 columns) of the strange nucleon correlators at pion mass $M_\pi \approx 700$~MeV from the a09m310, a12m220, a12m310 and a15m310 ensembles.
The gray band shown on each plot is the extracted ground-state matrix element from the two-sim fit that we use as our best value.
From left to right, the columns are:
the ratio of the three-point to two-point correlators with the reconstructed fit bands from the two-sim fit using the final $t_\text{sep}$ inputs, shown as functions of $t-t_\text{sep}/2$,
the one-state fit results for the three-point correlators at different  $t_\text{sep}$ values,
the two-sim fit results using $t_\text{sep}\in[t_\text{sep}^\text{min},t_\text{sep}^\text{max}]$ varying $t_\text{sep}^\text{min}$ and $t_\text{sep}^\text{max}$. }
\label{fig:SRatio-fitcomp}
\end{figure*}

On each lattice configuration, we calculate the nucleon two-point correlators using multiple sources:
\begin{equation}\label{}
C_N^\text{2pt}(P_z;t) =
 \langle 0|\Gamma\int d^3y\, e^{-iyP_z}\chi(\vec y,t)\chi(\vec 0,0)|0\rangle,
\end{equation}
with the nucleon interpolation operator $\chi$ as  $\epsilon^{lmn}[{u(y)^l}^Ti\gamma_4\gamma_2\gamma_5 d^m(y)]u^n(y)$ (where $\{l,m,n\}$ are color indices, $u(y)$ and $d(y)$ are quark fields),
the projection operator $\Gamma=\frac{1}{2}(1+\gamma_4)$,
$t$ is lattice Euclidean time,
and $P_z$ is the nucleon boost momentum along the spatial $z$-direction.
We use Gaussian momentum smearing~\cite{Bali:2016lva} on the quark field to improve the signal for nucleon boost momenta up to 3.0~GeV. 
Hundreds of thousands of measurements are made, varying for different ensembles.
Compared to our previous nucleon gluon PDF calculation on one a12m310 ensemble with $10^5$ measurements~\cite{Fan:2020cpa}, this study uses more measurements and varies the lattice spacing. 
We then calculate the three-point gluon correlator by combining the gluon loop with nucleon two-point correlators,
\begin{multline}\label{eq:3ptC}
 C_N^\text{3pt}(z,P_z;t_\text{sep},t) = \\
 \langle 0|\Gamma\int d^3y\, e^{-iyP_z}\chi(\vec y,t_\text{sep}){\cal O}_g(z,t)\chi(\vec 0,0)|0\rangle,
\end{multline}
where $t$ is the gluon-operator insertion time, $t_\text{sep}$ is the source-sink time separation.
${\cal O}_g(z,t)$ is the gluon operator introduced in Ref.~\cite{Balitsky:2019krf}:
\begin{equation}\label{eq:gluon_operator}
 {\cal O}(z)\equiv\sum_{i\neq z,t}{\cal O}(F^{ti},F^{ti};z)-\frac{1}{4}\sum_{i,j\neq z,t}{\cal O}(F^{ij},F^{ij};z),
\end{equation}
where the operator ${\cal O}(F^{\mu\nu}, F^{\alpha\beta};z) = F^\mu_\nu(z)U(z,0)F^{\alpha}_{\beta}(0)$, and $z$ is the Wilson link length. 
To extract the ground-state matrix element, we use a two-state fit on the two-point correlators and a two-sim fit on the three-point correlators:
\begin{multline}
C_N^\text{2pt}(P_z,t) = \\
|A_{N,0}|^2 e^{-E_{N,0}t} + |A_{N,1}|^2 e^{-E_{N,1}t} + \ldots,
\label{eq:2pt_fit_formula}
\end{multline}
\begin{align}
C_N^\text{3pt}&(z,P_z,t,t_\text{sep}) = \\\nonumber
 &|A_{N,0}|^2\langle 0|O_g|0\rangle e^{-E_{N,0}t_\text{sep}} \\\nonumber
{}+{} &|A_{N,0}||A_{N,1}|\langle 0|O_g|1\rangle e^{-E_{N,1}(t_\text{sep}-t)}e^{-E_{N,0}t} \\\nonumber
{}+{} &|A_{N,0}||A_{N,1}|\langle 1|O_g|0\rangle e^{-E_{N,0}(t_\text{sep}-t)}e^{-E_{N,1}t} \\\nonumber
{}+{} &|A_{N,1}|^2\langle 1|O_g|1\rangle e^{-E_{N,1}t_\text{sep}} \\\nonumber
{}+{} &\ldots,
\label{eq:3pt_fit_formula}
\end{align}
where the $|A_{N,i}|^2$ and $E_{N,i}$ are the ground-state ($i=0$) and first excited state ($i=1$) amplitude and energy, respectively.

To visualize our fitted matrix-element extraction, we compare to ratios of the three-point to the two-point correlator
\begin{equation}\label{eq:ratio}
R_N(z,P_z,t_\text{sep},t)=\frac{C_N^\text{3pt}(z,P_z, t, t_\text{sep})}{C_N^\text{2pt}(P_z,t)}.
\end{equation}
The left-hand side of Fig.~\ref{fig:LRatio-fitcomp} shows example ratios for the gluon matrix elements from all four ensembles at pion masses $M_\pi \in \{220,310\}$~MeV at selected momenta $P_z$ and Wilson-line length $z$.
The left column shows the ratio plots with data points of $R$ from different source-sink separation, $t_\text{sep}$, along with the reconstructed bands from the fit, showing how well the fit describing the data in Eq.~\ref{eq:ratio};
the final ground-state matrix elements are shown in grey bands.
We observe that the ratios increase with increasing source-sink separation $t_\text{sep}$ and continuously to approach the ground-state matrix elements 
obtained from the simultaneous two-state fit to three-point correlators with five inputs of $t_\text{sep}$. 
The middle and right columns of Fig.~\ref{fig:LRatio-fitcomp} show how the ground-state matrix elements vary with $t_\text{sep}^\text{min}$ and $t_\text{sep}^\text{max}$ with the same $t_\text{sep}^\text{max}$ and $t_\text{sep}^\text{min}$ used in the left column, respectively. 
The grey bands in each plot of the middle and right columns are the same ground-state matrix elements extracted from the fit shown in the corresponding plots in left column;
these demonstrate how stable our ground-state matrix element extractions are. 
We observe that overall the ground-state matrix elements are consistent with each other within one standard deviation with different $t_\text{sep}^\text{min}$ choices.
There seems to be some hint of $t_\text{sep}^\text{max}$-dependence in the ground-state matrix elements, but they do converge.
Taking the a12m220 ensemble as an example, we observe larger fluctuations in the matrix-element extractions when small $t_\text{sep}^\text{min}=3$, or
small $t_\text{sep}^\text{max}=6$ and 7, are used.
The ground-state matrix element extracted from two-sim fits comes into reasonable agreement when $t_\text{sep}^\text{min}>5$ and $t_\text{sep}^\text{max}>8$.
Similar results are observed with the nucleon when calculated at strange-quark mass, as shown in Fig.~\ref{fig:SRatio-fitcomp}.

\section{Results and Discussions}
\label{sec:results}

\subsection{Lattice-Spacing Dependence of RpITDs}

Using the nucleon ground-state matrix elements, we can now compute the reduced Ioffe-time pseudo-distribution (RpITD)~\cite{Radyushkin:2017cyf,Orginos:2017kos,Zhang:2018diq,Li:2018tpe}
\begin{equation}
\mathscr{M}(\nu,z^2)=\frac{\mathcal{M}(zP_z,z^2)/\mathcal{M}(0\cdot P_z,0)}{\mathcal{M}(z\cdot 0,z^2)/\mathcal{M}(0\cdot 0,0)},
\label{eq:RITD}
\end{equation}
where Ioffe time $\nu=zP_z$, and $\mathcal{M}(\nu,z^2)$ are the nucleon matrix elements at boost momentum $P_z$ and gluon operators with Wilson displacement $z$.
By construction, the renormalization of ${\cal O}(z)$ and kinematic factors are cancelled in the RpITDs, and the ultraviolet divergences are removed.
The RpITD double ratios employed here are normalized to one at $z=0$, and the lattice systematics are reduced due to the double ratio.
These RpITDs will be input into the pseudo-PDF framework detailed in Ref.~\cite{Radyushkin:2017cyf} to obtain the unpolarized nucleon gluon PDFs.

\begin{figure}[htbp]
  \centering
\includegraphics[width=0.45\textwidth]{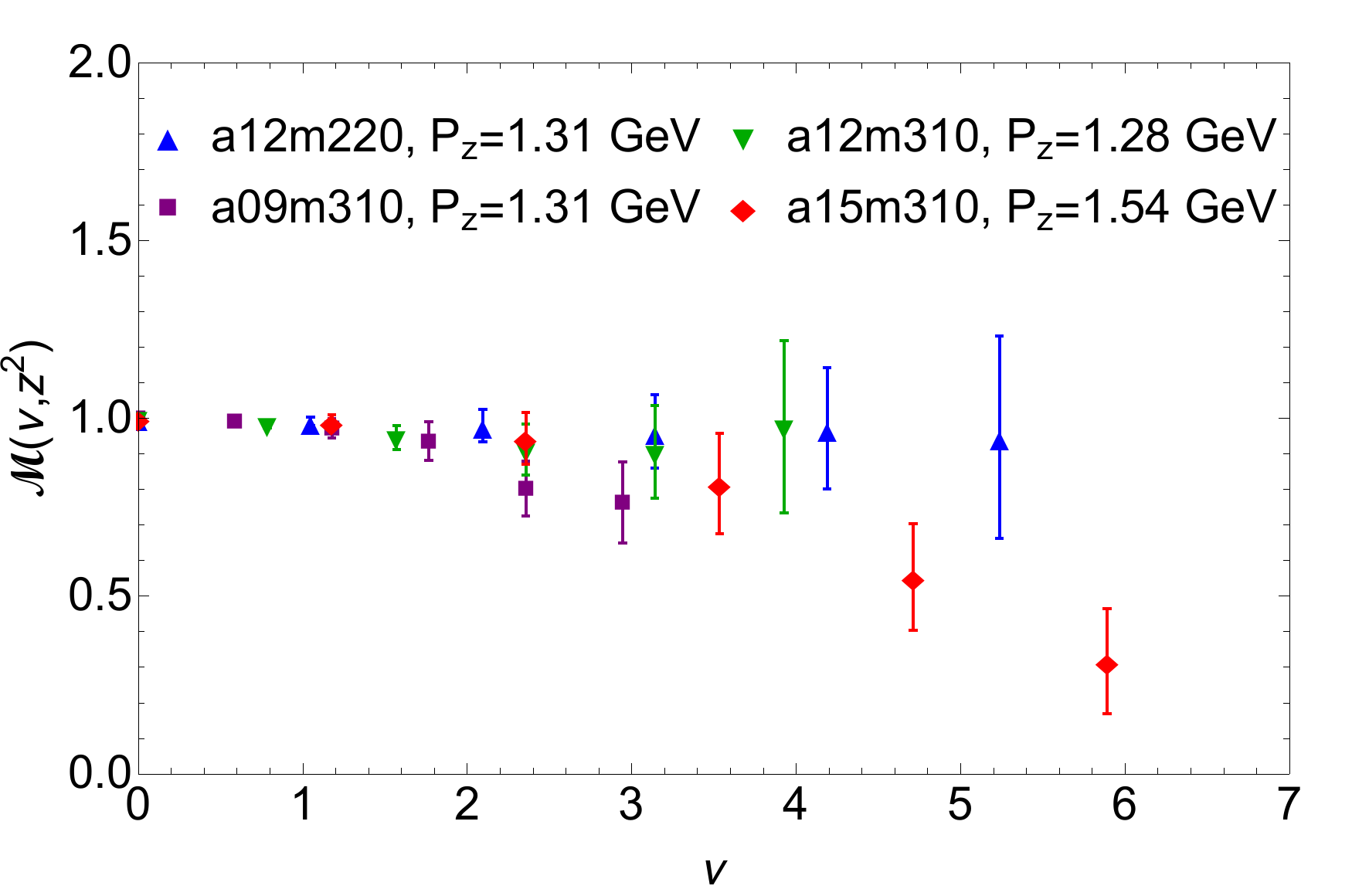}
\includegraphics[width=0.45\textwidth]{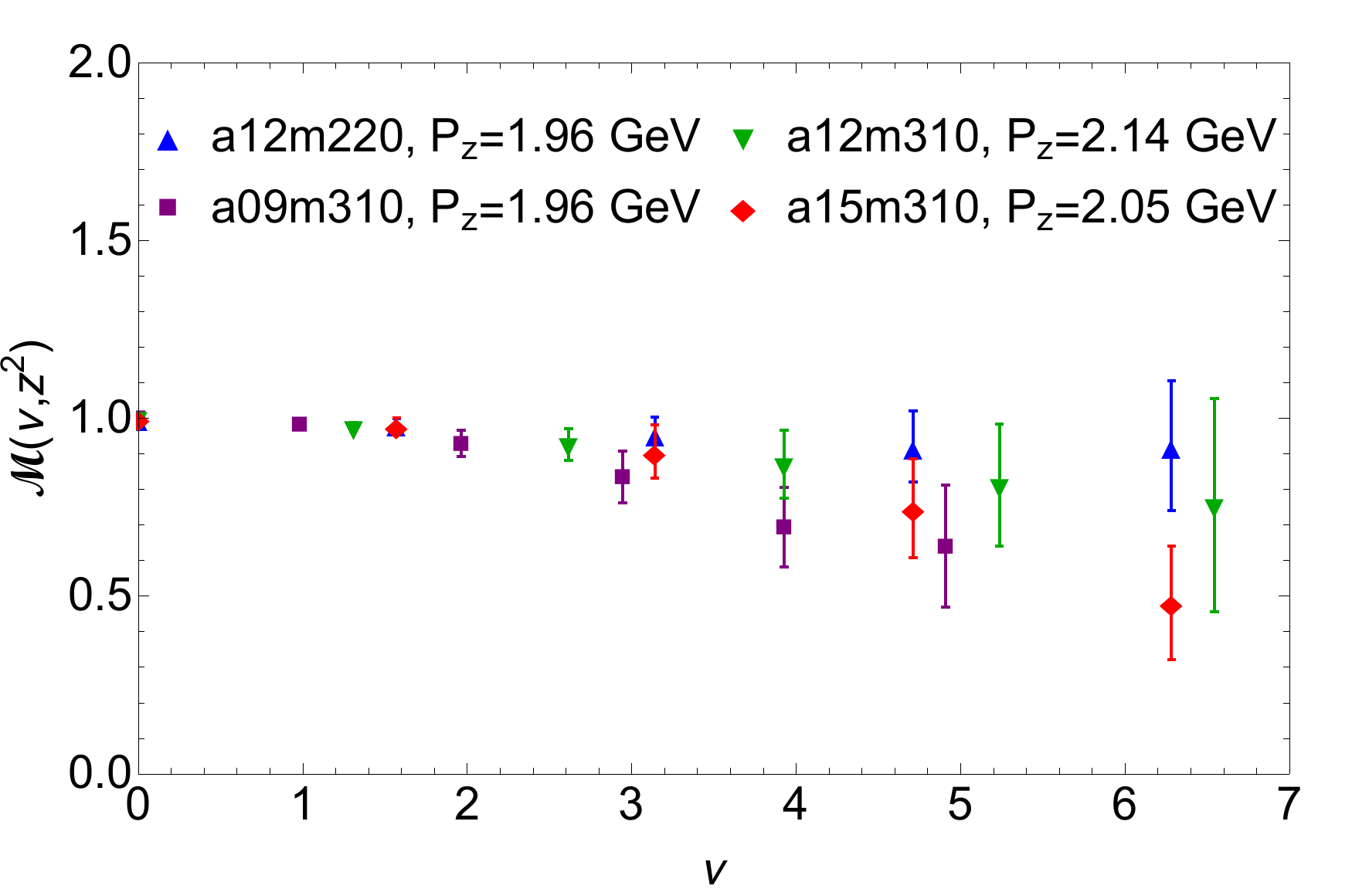}\\
\caption{ 
The RpITDs at boost momenta $P_z \approx 1.3$~GeV (top) and 2~GeV (bottom) as functions of Ioffe-time $\nu$ obtained from the fitted bare ground-state matrix elements for $M_\pi \approx \{220, 310, 310, 310\}$~MeV on a12m220, a09m310, a12m310, a15m310 ensembles, respectively.
There is no visible lattice-spacing or pion-mass dependence.
}\label{fig:RpITD-ensembles}
\end{figure}

We first examine the pion-mass and lattice-spacing dependence of the nucleon gluon RpITDs.
The top panel of Fig.~\ref{fig:RpITD-ensembles} shows the RpITDs at boost momentum around 1.3~GeV as functions of Ioffe time $\nu$ for the a12m220, a09m310, a12m310, and a15m310 ensembles.
Note that given the different size of the lattice dimensions and lattice spacings, there is no easy way to use the same boost momentum for all four ensembles.
For the a12m310, a12m220, and a09m310 ensembles, we are able to find close momenta, but the boost momentum on a15m310 is not so close, 1.54~GeV. 
Nevertheless, this allows us to study the lattice-spacing and pion-mass dependence without any interpolation; we find in both cases mild dependence.
The $a=0.12$~fm ensembles seem to prefer slightly higher central values for the RpITDs, but they are consistent with those from a09m310 and a15m310 within one standard deviation.

\begin{figure*}[htbp]
  \centering
\includegraphics[width=0.32\textwidth]{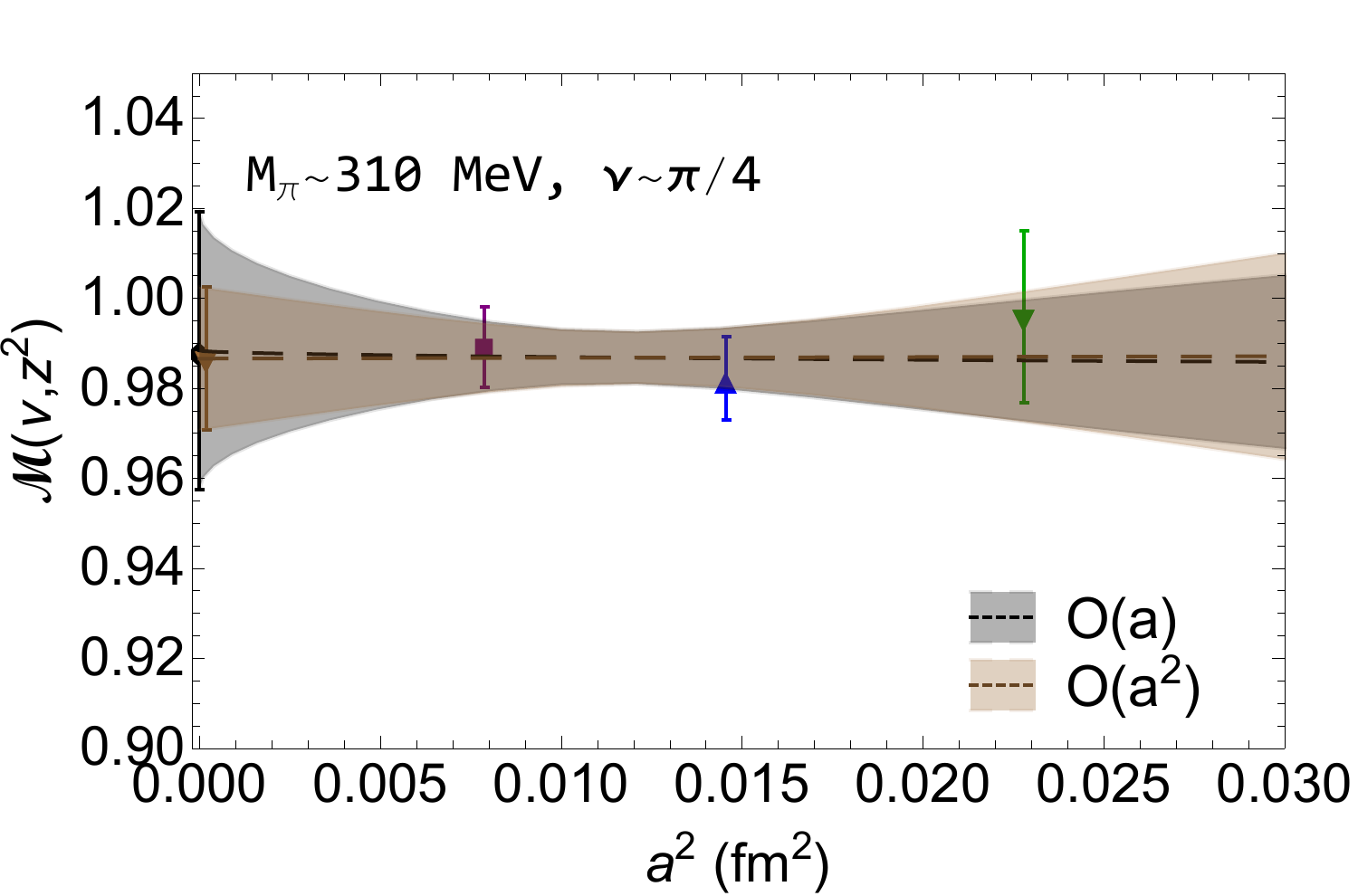}
\includegraphics[width=0.32\textwidth]{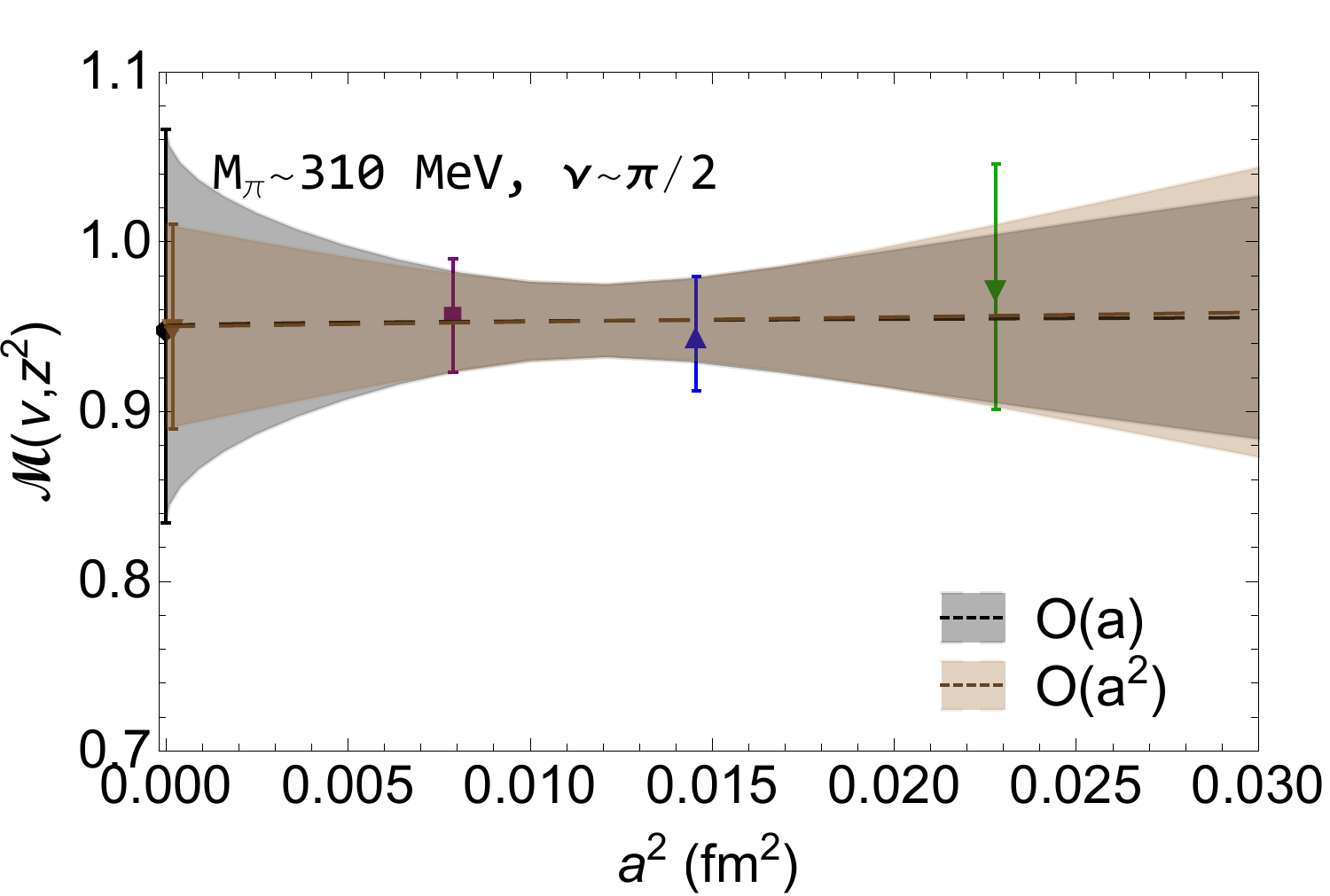}
\includegraphics[width=0.32\textwidth]{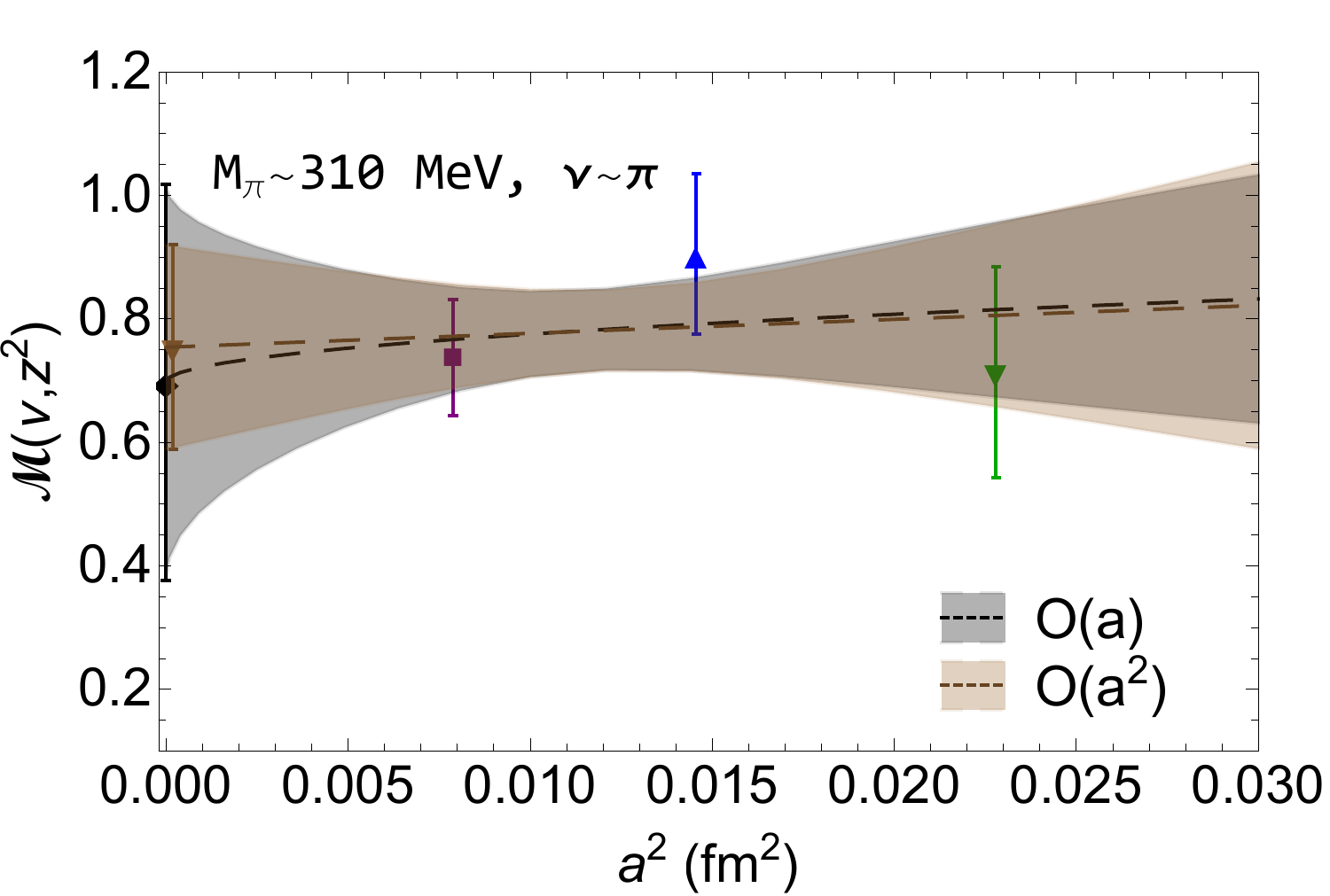}
\includegraphics[width=0.32\textwidth]{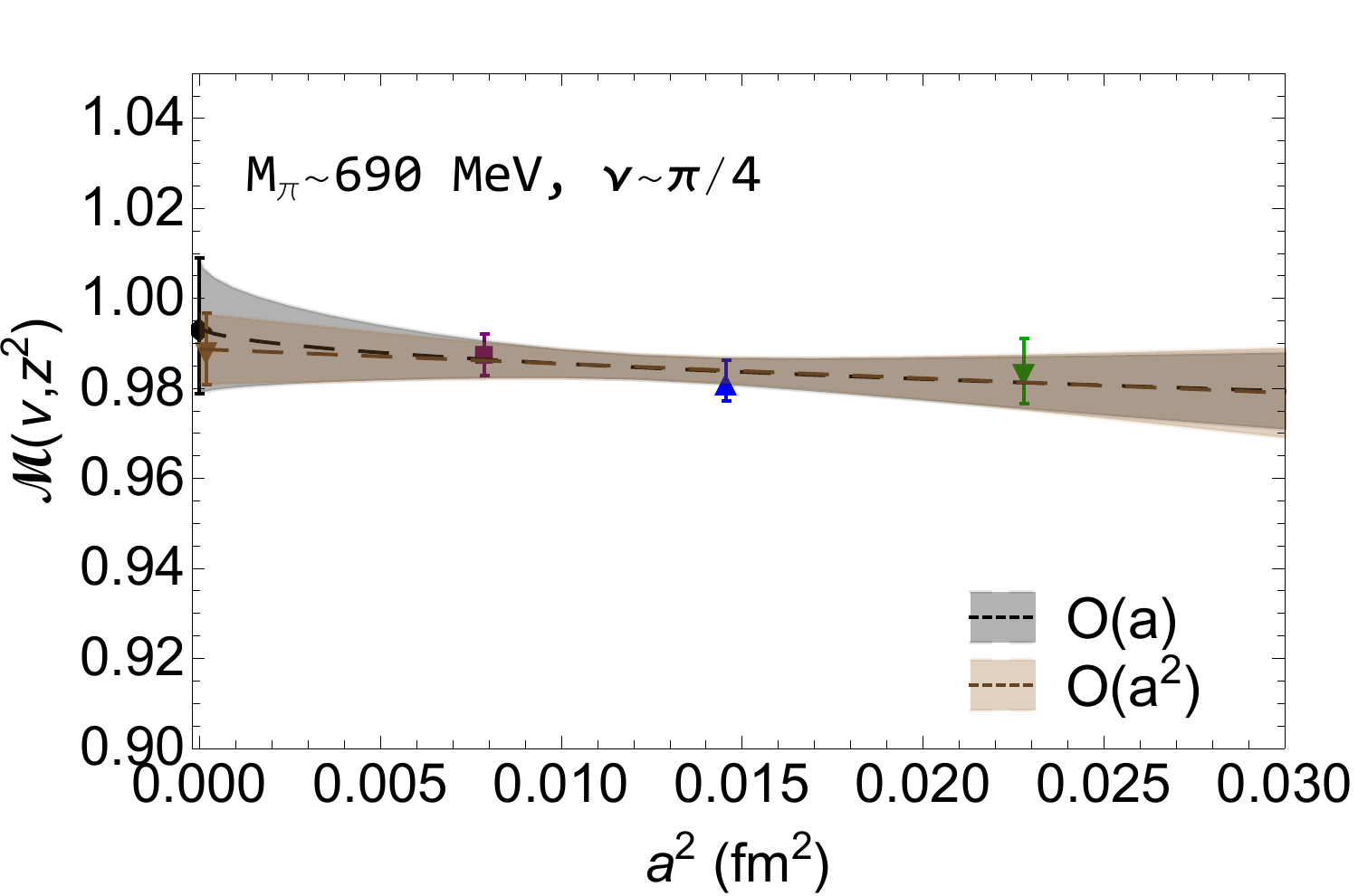}
\includegraphics[width=0.32\textwidth]{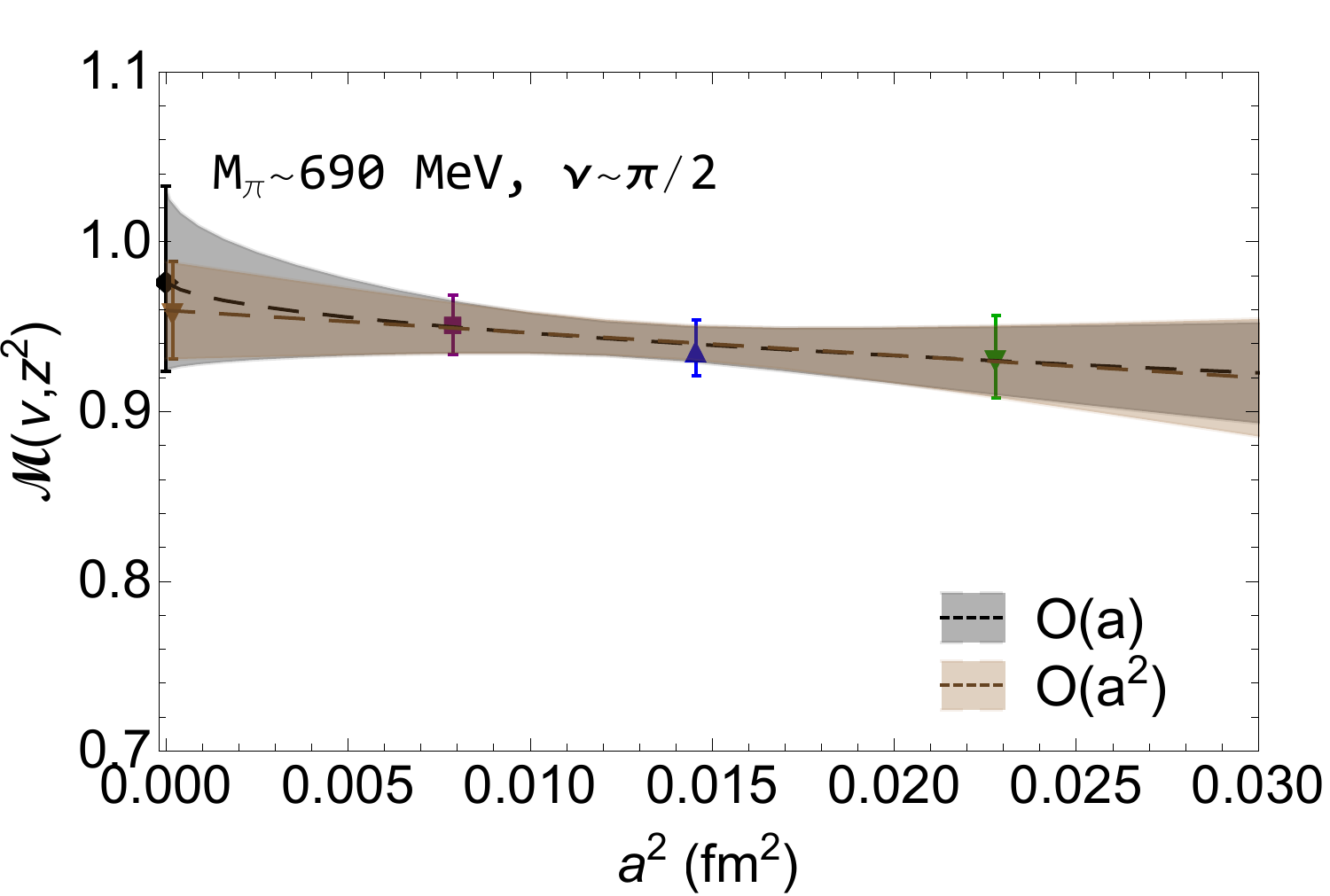}
\includegraphics[width=0.32\textwidth]{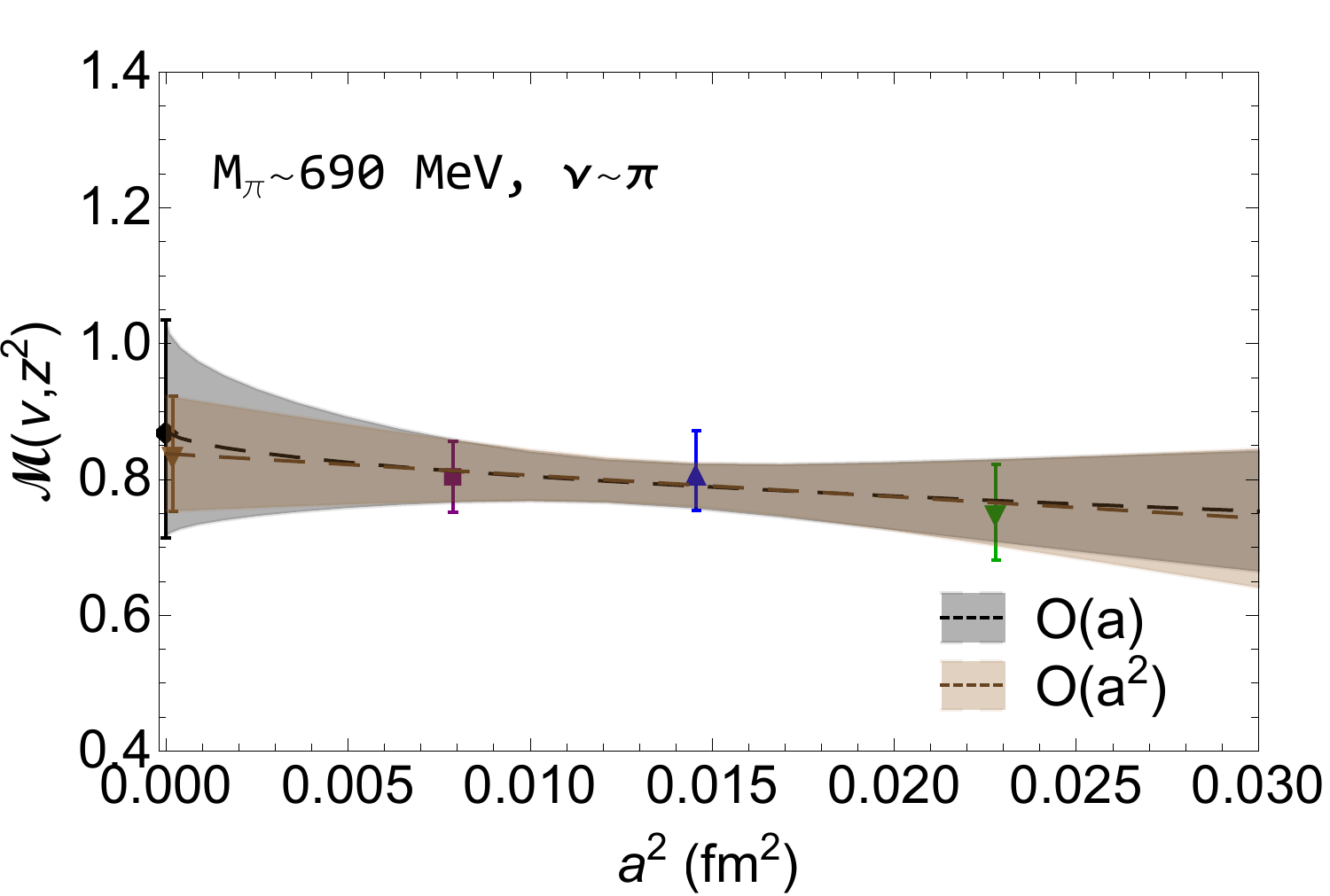}
\caption{
\label{fig:RpITD-a2}
RpITDs and the reconstructed bands from $O(a)$ or $O(a^2)$ fits of three lattice spacings for a09m310, a12m310 and a15m310 ensembles from left to right panel respectively, for both light (top  row) and strange nucleons (bottom row) at $\nu=\{\pi/4,\pi/2,\pi\}$.
}
\end{figure*}

We then choose a fixed pion mass and Ioffe time $\nu$ to explore the lattice-spacing dependence in more detail.
We examine the RpITDs at three values of $z$ and set $n_z \in \{4,3,2\}$ in lattice units ($P_z=\frac{2\pi}{L} n_z$) for the a09m310, a12m310 and a15m310 ensembles, respectively, to check the lattice-spacing dependence, where the Ioffe time $\nu=zP_z$ are the same for all ensembles.
We assume a linear $O(a)$ or $O(a^2)$ dependence and fit the RpITD data:
\begin{equation}
\mathscr{M}(\nu,z^2,a,M_\pi)= \mathscr{M}^\text{cont} + c_a a^n
\label{eq:RITD-a-a2}
\end{equation}
for $n=1$ and $2$, respectively. 
The fit results are shown in Fig.~\ref{fig:RpITD-a2} for both light and strange nucleons at $\nu=\{\pi/4,\pi/2,\pi\}$.
All the plots show the RpITD points at a fixed $\nu$ are consistent within one-sigma error;
however, there seems to be a trend that the lattice-spacing dependence is stronger at larger $\nu$, making the continuum-extrapolated results larger. 
The $\chi^2/\text{dof}$ are all within one standard deviation of 0.5, showing that there is no preference in choice of $\nu$ in this respect.
In both $O(a)$ and $O(a^2)$ extrapolation, the light and strange nucleons appear to have opposite trends with lattice spacing; however, the slopes of the fits are consistent with zero within one standard deviation for each pion mass over all the selected $\nu$.
Corresponding to a reading of Fig.~\ref{fig:RpITD-a2} left to right, top to bottom, the $O(a)$ $\mathscr{M}^{\text{cont}}$ are 0.988(31), 0.95(12), 0.70(32), 0.994(15), 0.978(54), and 0.87(16). The $O(a^2)$ values are 0.987(16), 0.950(60), 0.75(16), 0.9997(80), 0.959(29), and 0.837(84). 
The deviation of the continuum-limit RpITD increases as $\nu$ increases, with the continuum-limit error at $\nu = \pi$ consistently being about a factor of ten larger than those at $\nu = \pi/4$.
The continuum-limit RpITDs using $a$ and $a^2$ extrapolation are consistent with each other, but the former has larger error, due to extrapolating over a larger distance.
The error in the light-nucleon extrapolated RpITD values is consistently double that of the strange nucleon. 

\subsection{Continuum-Physical Extrapolation}

\begin{figure}[htbp]
\centering
\centering
\includegraphics[width=0.46\textwidth]{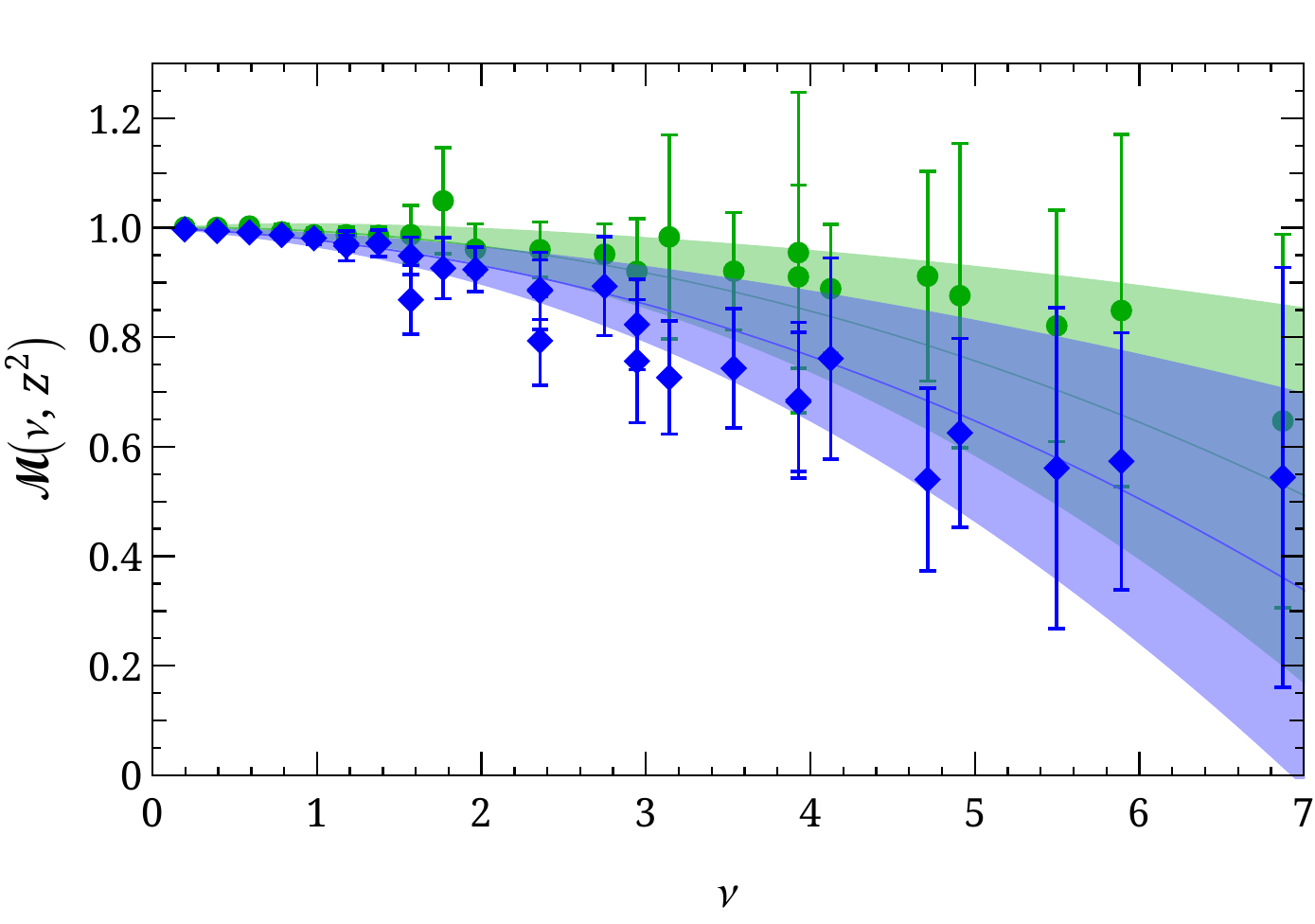}
\includegraphics[width=0.46\textwidth]{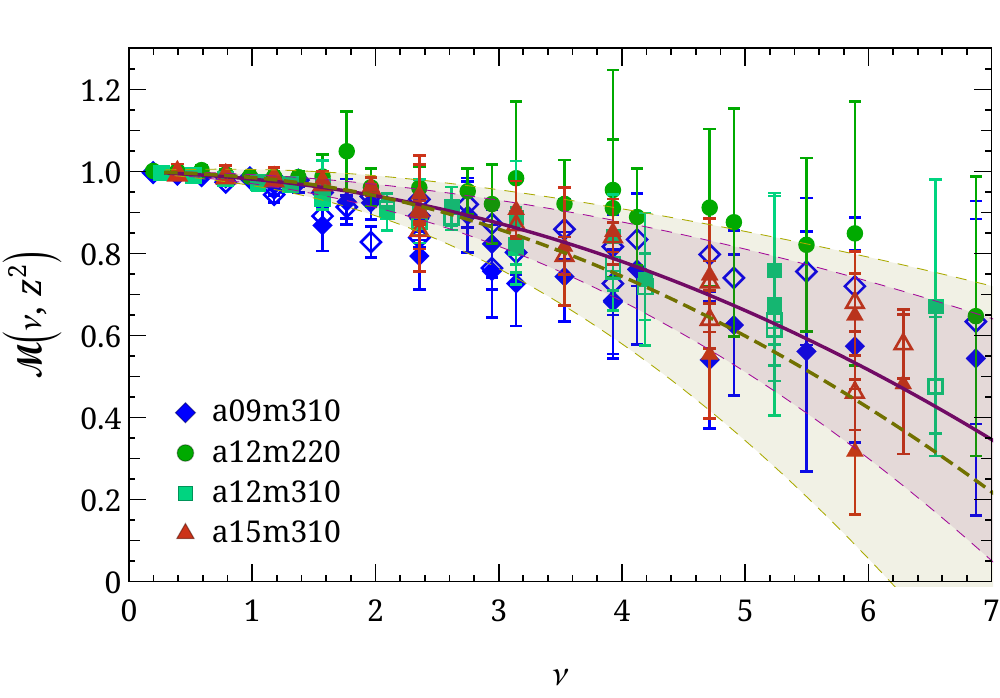}
\caption{
(Top) Examples of the RpITDs $\mathscr{M}$ reconstructed bands from fits in Eq.~\ref{eq:RpITD-extra} for a09m310 (blue points and light blue band), a12m220 (green) lattice ensembles.
The fit ansatz is able to describe the data well.
(Bottom)
Collected data for all ensembles with $a$ (dashed band) and $a^2$ (solid band) continuum extrapolation at the physical pion mass.
Open symbols indicates the data point from the same-symbol ensemble but at the heavier quark mass.
}
\label{fig:RpITD_exta_1}
\end{figure}

We extrapolate the nucleon gluon PDF at the physical pion mass and continuum limit based on the RpITDs for the a12m220, a09m310, a12m310 and a15m310 ensembles.
Given that we observe small differences between $a$ and $a^2$ extrapolation at fixed $M_\pi$ and $\nu$ in the previous subsection, we focus on $a^2$ extrapolation here.
We use the following ansatz, linear in pion mass and lattice-spacing squared, to extrapolate to the continuum-physical limit:
\begin{align}\label{eq:RpITD-extra}
\mathscr{M}(\nu, z^2,a,M_{\pi})={}&\left(\sum_{k=0}^{k_\text{max}}\lambda_k(a,M_{\pi}) \nu^k+c_z(a,M_{\pi})z^2\right) \nonumber \\
{}\times{}&(1+c_a a^2+c_M(M_\pi^2-(M_\pi^{\text{phys}})^2),
\end{align}
Given the noisiness of the gluon RpITD data, $k_\text{max}=2$ is used in this study.
We found that in all the ensembles in our calculation, the $z^2$-dependence, $c_z(a,M_{\pi})$ is consistent with zero within two standard deviations.
The example reconstructed fitted bands for a09m310 and a12m220 are shown in the top plot of Fig.~\ref{fig:RpITD_exta_1}.
We drop data points at a $\nu$ value if they have errors more than twice as large another data point at that $\nu$;
this improves the clarity of the diagram showing the extrapolation effects.
Our fit ansatz is able to describe our data reasonably with $\nu$ up to about 7;
in the future, when larger $P_z$ is used to reach larger $\nu$, a better interpolation form will likely be needed to describe the small- and large-$\nu$ regions well.
We show the continuum-physical RpITD band on the top plot of Fig.~\ref{fig:RpITD_exta_1} with all the data points from the four ensembles and $a^2$ extrapolation to the continuum-physical band.
The open symbols indicate the strange-mass nucleon calculation from the ensemble.
With the a15m310, a12m310 and a09m310 ensembles, since we have the same number of measurements for both strange and light nucleons, within each ensemble we bootstrap the light and strange renormalized matrix elements to keep the correlations.
Across the ensembles, the data are independent and the typical bootstrap treatment is used.
For comparison purposes, we also replace the $a^2$ term in Eq.~\ref{eq:RpITD-extra} with $a$;
its physical-continuum results are shown as the band with dashed center line in Fig.~\ref{fig:RpITD_exta_1}. 
Similar to what we observe in Fig.~\ref{fig:RpITD-a2}, both extrapolations give us a consistent continuum-physical limit (within one standard deviation) with Ioffe time $\nu$ up to 7, but the $O(a)$-extrapolated continuum-physical RpITD has larger errors, especially in the larger-$\nu$ region.

\subsection{Gluon PDF Results}

With the physical-continuum RpITD obtained in the previous section, we can now extract the gluon PDF distribution using the pseudo-PDF matching condition~\cite{Balitsky:2019krf} that connects the RpITD $\mathscr{M}$ to the lightcone gluon PDF $g(x,\mu^2)$:
\begin{equation}
\label{eq:matching-gg}
\mathscr{M}(\nu,z^2) = \int_0^1 dx \frac{xg(x,\mu^2)}{\langle x \rangle_g}R_{gg}(x\nu,z^2\mu^2),
\end{equation}
where $\mu$ is the renormalization scale in the $\overline{\text{MS}}$ scheme
and $\langle x \rangle_g=\int_0^1 dx \, x g(x,\mu^2)$ is the gluon momentum fraction of the nucleon.
$R_{gg}$ is the gluon-in-gluon matching kernel originally derived in Ref.~\cite{Balitsky:2019krf} and has been used in previous gluon PDF lattice works~\cite{Fan:2020cpa,Fan:2021bcr,Salas-Chavira:2021wui,HadStruc:2021wmh}.
We use the RpITD extrapolated using the $a^2$ term (in Eq.~\ref{eq:chi2-RITD-fit}) at physical pion mass and continuum limit with a fit range of $\nu\in[0,7]$, corresponding to the region where we have data in all ensembles.
We ignore the quark-PDF contribution to the RpITDs in this calculation;
it is likely to be small, based on our past study of the pion gluon PDF~\cite{Fan:2021bcr}.
We will later estimate the quark contribution as systematic effect.
One can obtain the gluon PDF $g(x,\mu^2)$ by fitting the RpITD through the matching condition in Eq.~\ref{eq:matching-gg}. 
We adopt the phenomenologically motivated form commonly used in the global analysis
\begin{equation}
f_g(x,\mu) = \frac{xg(x, \mu)}{\langle x \rangle_g(\mu)} = \frac{x^A(1-x)^C}{B(A+1,C+1)},
\label{functional}
\end{equation}
for $x\in[0,1]$ and zero elsewhere.
The beta function $B(A+1,C+1)=\int_0^1 dx\, x^A(1-x)^C$ is used to normalize the area to unity.
Such a form is also used in global fits to obtain the global PDF fits, such as nucleon gluon PDF by CT18~\cite{Hou:2019efy} 
and the nucleon and pion gluon PDF by JAM~\cite{Moffat:2021dji,Barry:2018ort,Cao:2021aci}. 
We then fit the lattice physical-continuum RpITDs $\mathscr{M}^\text{lat} (\nu,z^2,a,M_\pi)$ obtained in Eq.~\ref{eq:RpITD-extra} to the parametrization form $\mathscr{M}^\text{fit}(\nu,\mu,z^2,a,M_\pi)$ in Eq.~\ref{eq:matching-gg} by minimizing the $\chi^2$ function,
\begin{multline}
\label{eq:chi2-RITD-fit}
\chi^2(\mu,a,M_\pi) = \\
    \sum _{\nu,z} \frac{(\mathscr{M}^\text{fit}(\nu,\mu,z^2,a,M_\pi) - \mathscr{M}^\text{lat}(\nu,z^2,a,M_\pi))^2}{\sigma^2_{\mathscr{M}}(\nu,z^2,a,M_\pi)}.
\end{multline}

\begin{figure}[htbp]
\centering
\centering
\includegraphics[width=0.48\textwidth]{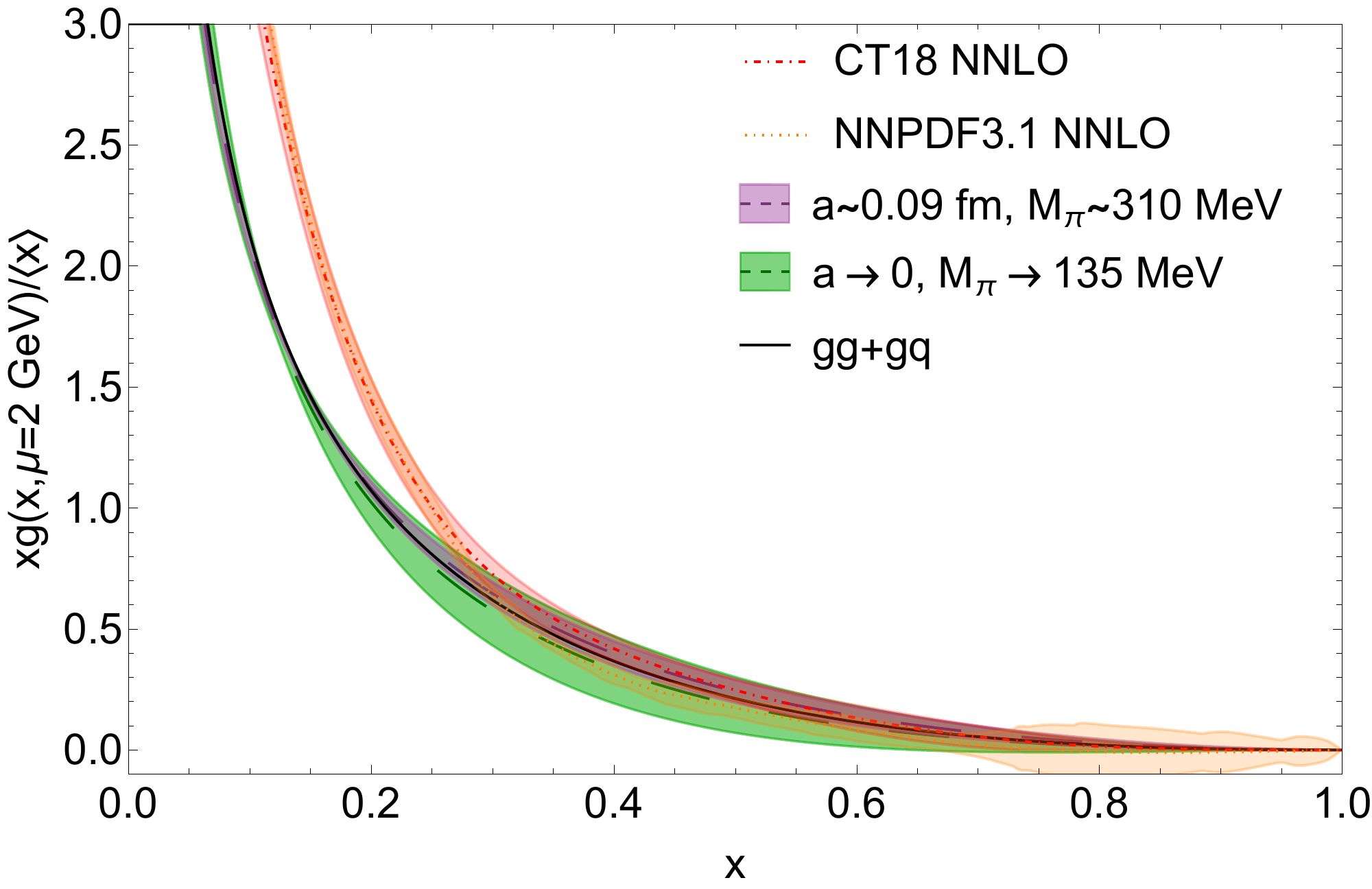}
\includegraphics[width=0.48\textwidth]{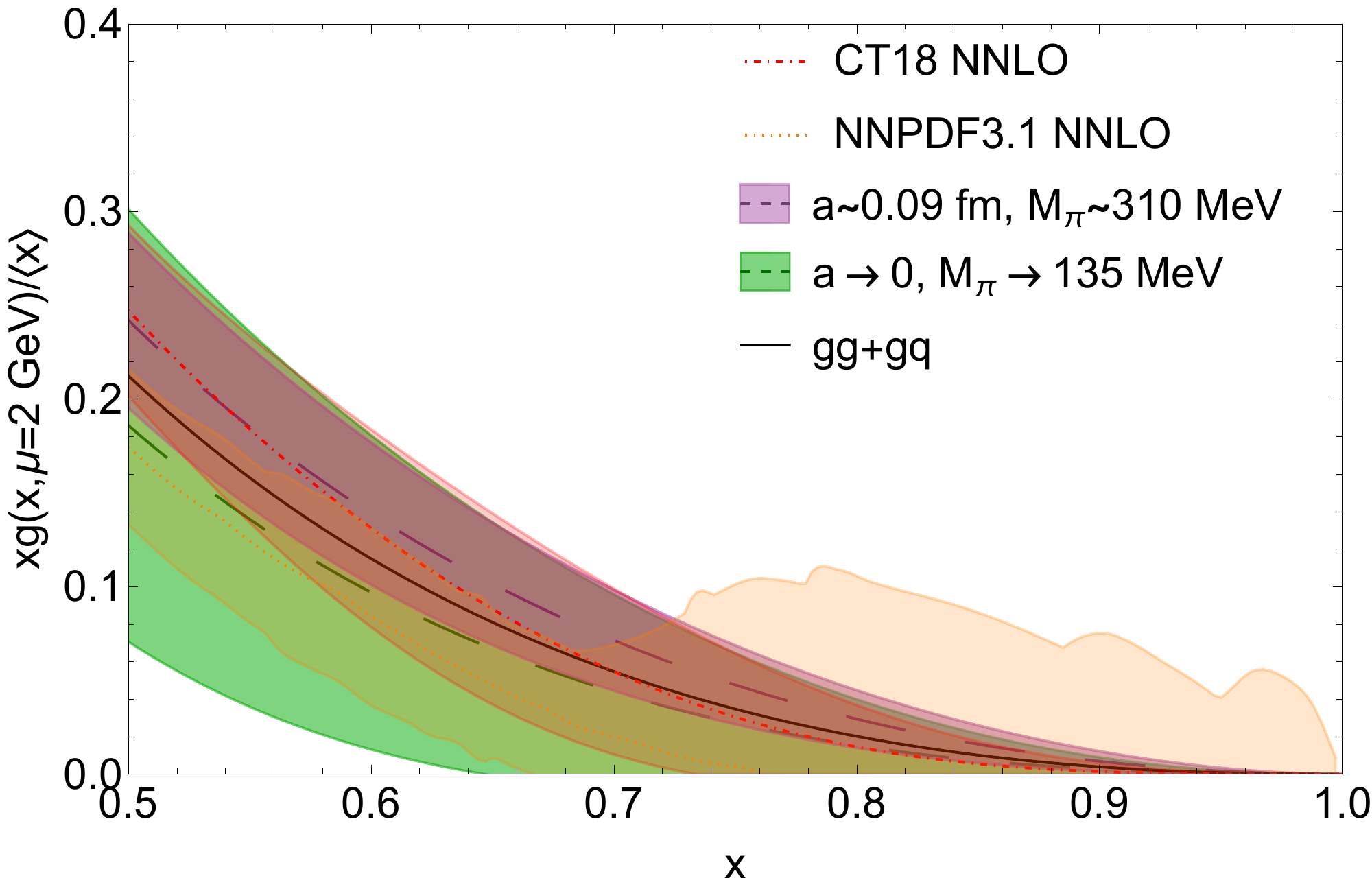}
\caption{
The unpolarized gluon PDF, $xg(x,\mu)/\langle x \rangle_g$ as a function of $x$ and its zoomed in plot, obtained from the fits to the smallest--lattice-spacing ensemble data compared with the fit to the data obtained from extrapolation to physical pion mass and continuum limit.
The black solid line is the central value of the fit to the continuum-physical PDFs, including the gluon-in-quark term in the matching, using CT18 for the quark PDF contributions.
The results from the global fits by CT18~\cite{Hou:2019efy} and NNPDF3.1~\cite{Ball:2017nwa} NNLO gluon PDFs are also shown in the plots, and our gluon PDF results are consistent with the global fits for $x \in [0.3,1]$
}
\label{fig:xgx-comp}
\end{figure}

Our results for the continuum-physical unpolarized gluon PDF $xg(x,\mu)/\langle x \rangle_g$ are shown in Fig.~\ref{fig:xgx-comp}, along with the same determination from the smallest lattice-spacing ensemble obtained in this work, and selected global-fit gluon PDFs from CT18~\cite{Hou:2019efy} and NNPDF3.1~\cite{Ball:2017nwa} NNLO analysis.
The gluon distribution in continuum-physical limit has much larger errors by a factor of 3--5 than those obtained from single--lattice-spacing analysis, due to the continuum extrapolation.
Overall, the results from single-ensemble calculations on a09m310 are consistent with the continuum-physical one (which has much larger uncertainties).
To reduce the errors in the continuum-physical distribution will be difficult, since it requires reduced errors in all ensembles, increasing the calculation cost by at least another order of magnitude.
Both of our lattice distributions agree with the global-fit gluon distribution at mid to large $x$ but deviate for $x<0.3$.
This is likely due to lack of large-$\nu$ lattice data in the input, which has higher sensitivity to the smaller-$x$ data.
Future calculations to push for even larger $P_z$ will be needed to improve the small-$x$ gluon distribution.

We now consider the systematic uncertainty coming from neglecting the contribution of the quark term, $\frac{P_z}{P_0}\int_0^1 dx \frac{xq_S(x,\mu^2)}{\langle x \rangle_g}R_{gq}(x\nu,z^2\mu^2)$ in Eq.~\ref{eq:matching-gg}.
We ignored this contribution initially based on the assumption (motivated by global fits) that the nucleon total quark PDF $q_S(x)$ is smaller than the gluon PDF. 
We can estimate the systematic due to omitting the $q_S(x)$  contribution by using the nucleon flavor-dependent quark PDFs from CT18 at NNLO~\cite{Hou:2019efy}.
Following a similar procedure to Ref.~\cite{Fan:2021bcr}, we add the quark-gluon mixing term to the extraction of $xg(x)/\langle x \rangle_g$ from the RpITD in Eq.~\ref{eq:matching-gg}.
The central value of the updated $xg(x)/\langle x \rangle_g$ including both gluon-in-gluon (gg) and gluon-in-quark (gq) contributions is shown in Fig.~\ref{fig:xgx-comp} (black solid line).
The difference is much smaller than the current statistical errors, so we will ignore this quark contribution in this calculation.
However, in the future, when the continuum-physical gluon PDF precision are improved, we should re-examine this contribution more carefully.

\begin{figure}[htbp]
\centering
\centering
\includegraphics[width=0.48\textwidth]{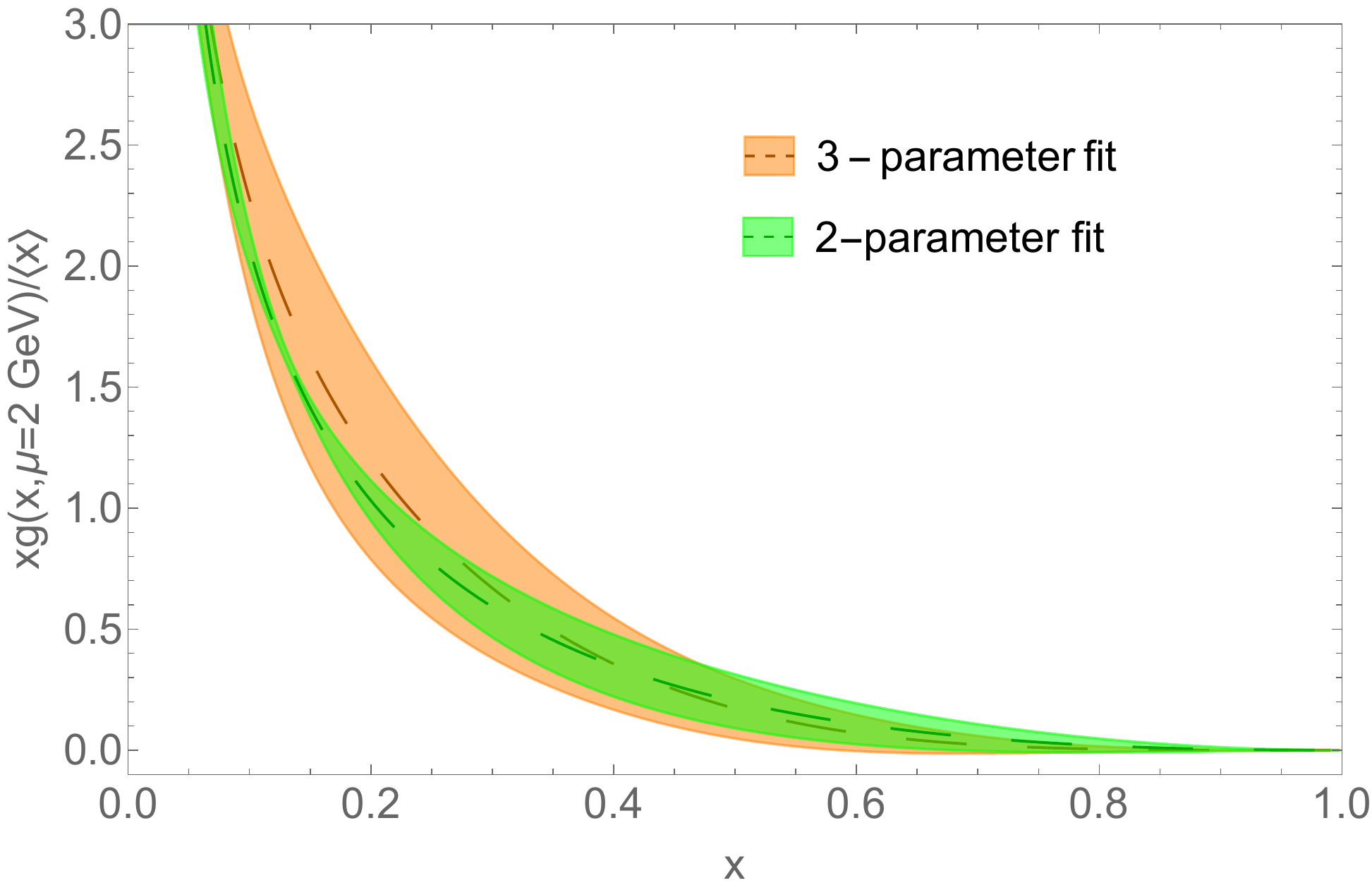}
\caption{
$xg(x, \mu)/\langle x \rangle_g$ at $\mu^2=4\text{ GeV}^2$ as function of $x$, extracted from continuum-physical RpITDs using the two- (green) and three-parameter (orange) fits described in Eqs.~\ref{functional} and \ref{eq:functional3}, respectively.
The results are consistent within statistical errors.
}
\label{fig:xgx-fitform-comp}
\end{figure}

We investigate the systematic uncertainty introduced by the choice of parametrization form used for $f_g(x,\mu)$.
We consider a three-parameter form used in PDF global analysis and some lattice calculations,
\begin{equation}
f_{g,3}(x,\mu) = \frac{x^A(1-x)^C(1+D {x})}{B(A+1,C+1)+DB(A+2,C+1)}.
\label{eq:functional3}
\end{equation}
We fit our continuum-physical--limit RpITDs to this form up to maximal Ioffe time $\nu_\text{max}=7$.
A comparison of the fit-form choice is shown in Fig.~\ref{fig:xgx-fitform-comp}.
We find the goodness-of-fit improves slightly due to the introduction of a new free parameter in the fit form, but the gluon PDF results are noisier and are consistent with the two-parameter fit.
We will use $xg(x, \mu)/\langle x \rangle_g$ from the two-parameter fit as our main result for this work.


\begin{figure}[t!]
\centering
\includegraphics[width=0.48\textwidth]{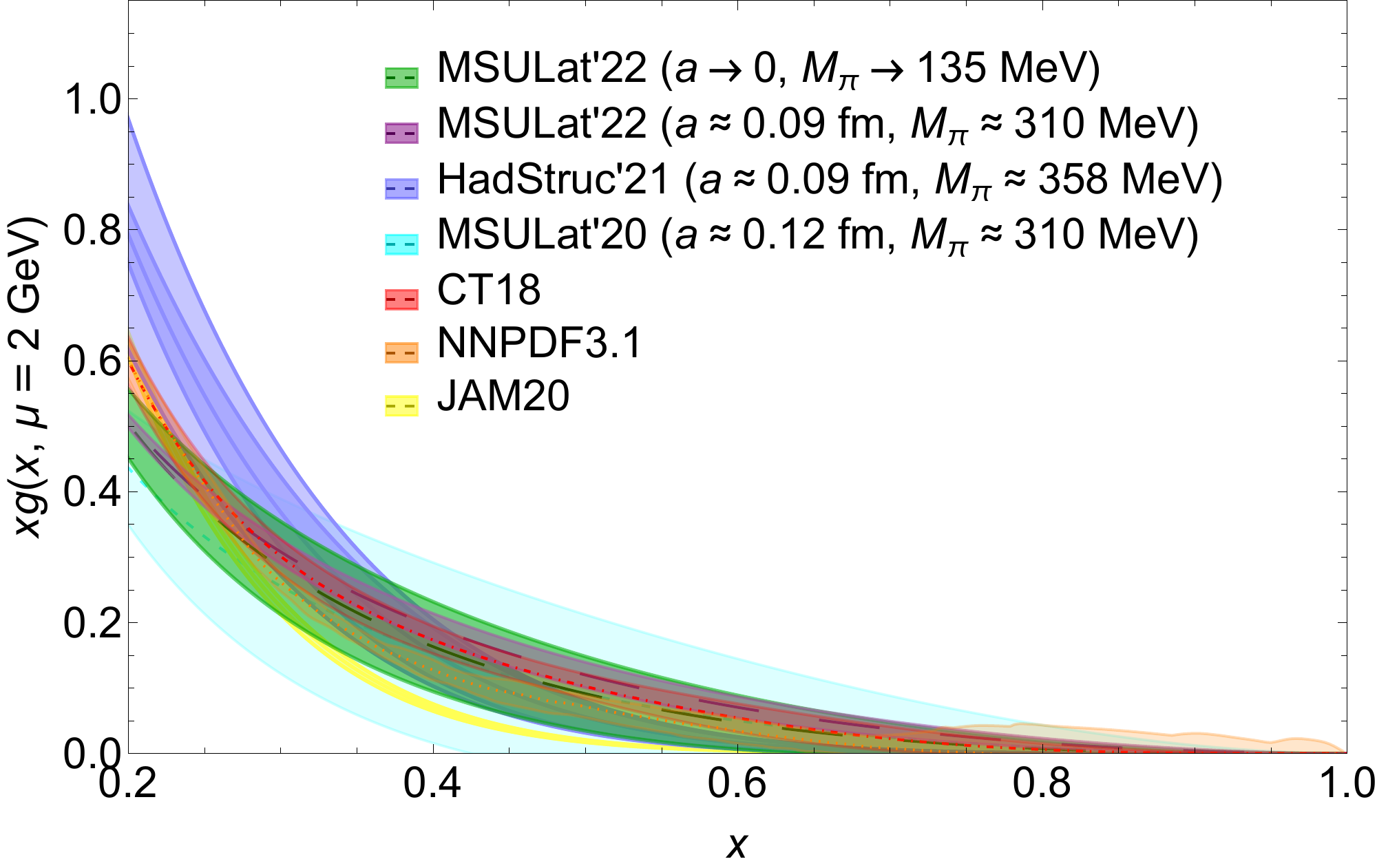}
\includegraphics[width=0.48\textwidth]{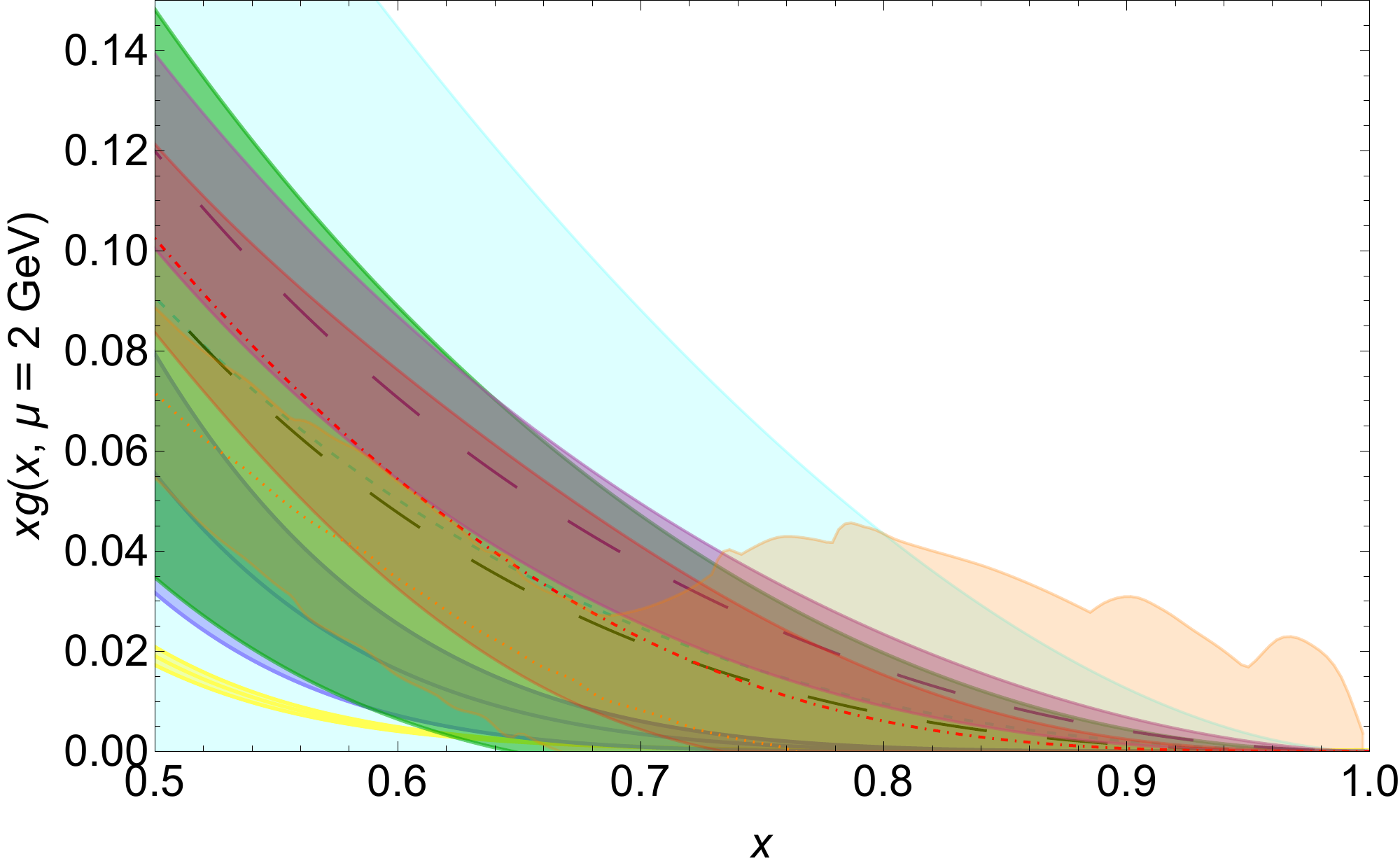}
\caption{
The unpolarized gluon PDF, $xg(x,\mu)$ a function of $x$ with $x \in [0.2,1]$ (top) region as and a close-look of the large-$x$ region (bottom), obtained from our continuum-physical (green) and a09m310-ensemble (purple) RpITDs  compared with a single-ensemble analysis from HadStruc ($a\approx 0.094$~fm, $M_\pi \approx 358$~MeV), and the CT18 NNLO~\cite{Hou:2019efy} (red band),  NNPDF3.1 NNLO~\cite{Ball:2017nwa} (orange ban) and JAM20~\cite{Moffat:2021dji} (yellow band) gluon PDFs
 at $\mu=2$~GeV in the $\overline{\text{MS}}$ scheme.
Other prior lattice calculations of $xg(x)$ (including those done at single ensemble) from
HadStruc~\cite{HadStruc:2021wmh} (blue band) and MSULat~\cite{Fan:2020cpa} (cyan band) are also shown in the plot.
Our PDF results are consistent with the CT18 NNLO and NNPDF3.1 NNLO unpolarized gluon PDFs within errors.
}
\label{fig:xg-comp}
\end{figure}

The unpolarized nucleon gluon PDF $xg(x)$ can be extracted by taking the ratio of $f_g(x,\mu)=xg(x, \mu)/\langle x \rangle_g(\mu)$ and the gluon momentum fraction $\langle x \rangle_g(\mu)$ obtained in Ref.~\cite{Fan:2022qve}.
Reference~\cite{Fan:2022qve} calculated the gluon momentum fraction using valence clover fermion action on 0.09-, 0.12-, and 0.15-fm HISQ 2+1+1-flavor lattice ensembles with three pion masses, $220$, $310$ and $690$~MeV.
The renormalization was done using RI/MOM nonperturbative renormalization in $\overline{\text{MS}}$ scheme at 2~GeV and using cluster-decomposition error reduction (CDER) to enhance the signal-to-noise ratio of the renormalization constant~\cite{Liu:2017man,Yang:2018bft}.
The gluon momentum fraction was extrapolated to the continuum-physical limit and found to be consistent with other recent lattice-QCD results at physical pion mass. 
Our final unpolarized nucleon gluon PDF $xg(x)$ extrapolated to physical pion mass $M_\pi=135$~MeV and the continuum limit $a\to 0$ is shown as green bands in Fig.~\ref{fig:xg-comp};
once again, we found reasonable agreement with the global fits from CT18~\cite{Hou:2019efy} and NNPDF3.1~\cite{Ball:2017nwa} NNLO analysis for $x \in [0.25,1]$, even thought the the gluon momentum fraction obtained from the global fits is about two-sigmas lower than the lattice calculations. 
We do observe tension with gluon PDF from JAM20~\cite{Moffat:2021dji} analysis for $x <0.6$ regions but its gluon PDF also behave quite different from the the CT18 and NNPDF results, even with smaller errors; we look forward to updates on the global-fit community on resolving these discrepancy.

We also compare our results with other previous lattice-QCD calculation on $xg(x)$. The light cyan bands in Fig.~\ref{fig:xg-comp} shown the first pseudo-PDF calculation  done using clover-on-HISQ with 0.12-fm lattice spacing and 310- and 700-MeV pion mass using 898 lattice configuration  with 32 sources  per configuration for nucleon two-point correlators\cite{Fan:2020cpa}. The results are extrapolated to physical pion mass using naive two valence pion mass extrapolation with $xg(x)$  reconstructed by multiplying the gluon momentum fraction taken from Ref.~\cite{Lin:2020rut}. 
The blue bands in Fig.~\ref{fig:xg-comp} show a followup calculation performed by HadStruc collaboration using 2+1 dynamical flavors of clover fermions with stout-link smearing on the gauge fields, 0.09-fm lattice spacing, 358-MeV pion mass, and 64 source measurements on 349 lattice configurations with gradient-flow improved gluonic operators~\cite{HadStruc:2021wmh}.
They used multiple nucleon interpolating fields, allowing them to use generalized eigenvalue method to determine the best overlap with ground-state nucleon gluonic matrix elements.
They used the gluon momentum fraction obtained from an independent lattice work (2+1+1-flavor at physical pion mass) to determine $xg(x)$.
The outer blue bands indicate their uncertainty estimated from $\left\langle x \right\rangle_g$.
We also show our result on the ensemble with 0.09-fm lattice-spacing and 310-MeV pion mass as a purple band in Fig.~\ref{fig:xg-comp}; it used about 300k measurements spread out over 1000 lattice configurations.
Our single-ensemble results have errors comparable to (in some regions, smaller than) CT18 and NNPDF.
The lattice-spacing and pion-mass here is similar to those used in the HadStruc calculation~\cite{HadStruc:2021wmh} but without the additional uncertainties due to continuum-physical extrapolation (shown as a green band).
There are noticeable deviations from the HadStruc results, especially in the larger-$x$ region;
their large-$x$ gluon PDF is much smaller than ours.
However, given that multiple methodological aspects are done quite differently (for example, we used the momentum fraction from the same lattice ensemble and different gluon-operator smearing), it may require the full calculation, including continuum-physical extrapolation, to have meaningfully compare them. 
All the prior single-ensemble lattice results (without the systematics from lattice discretization) agree with our continuum-physical $xg(x)$ due to the larger total errors from the continuum-physical extrapolation.
Future work to include finer lattice-spacing and 220-MeV or lighter pion masses in the extrapolation will help to improve the continuum-physical determination of the lattice gluon PDF.

\section{Summary and Outlook}
\label{sec:summary}

We extracted the nucleon $x$-dependent gluon PDFs $xg(x)$ using clover fermions as valence action and 310-MeV 2+1+1 HISQ configurations generated by the MILC Collaboration at three pion masses and three lattice spacings.
We found their dependence to be weak at the current statistics of hundreds of thousands of nucleon two-point correlator measurements.
We removed the excited-state contributions to the ground-state matrix elements using a two-state fitting strategy and studied the stability of the extraction of the ground-state matrix elements with various fit ranges.
We then calculated the reduced pseudo-ITD using the fitted matrix elements and studied their pion-mass and lattice-spacing dependence, which are also mild.
We then extrapolated the reduced pseudo-ITDs to physical-continuum limit before   extracting the gluon parton distribution $xg(x)/\langle x \rangle_g$ in the $\overline{\text{MS}}$ scheme at 2~GeV.
Using the nonperturbatively renormalized nucleon momentum fraction calculated on clover-on-HISQ ensembles, we were able to compare our single-ensemble $xg(x)$ calculations at lattice spacing 0.09~fm with prior lattice calculations.
We found both our 0.09-fm and continuum-physical limit $xg(x)$ to be in good agreement with CT18 and NNPDF3.1 NNLO global-fit results in the range $x \in [0.2,1]$ in $\overline{\text{MS}}$ scheme at 2~GeV.
We have used the CT18 quark PDFs to estimate the quark-gluon mixing and checked the gluon PDF fit-form dependence, but these errors are not included in our continuum-physical results, since the statistical errors are large in comparison.
Future work should include finer lattice spacings, even higher statistics to improve the signal and larger boost momentum to expand the range of $\nu$, which will improve the reliability of the results at small $x$.

\section*{Acknowledgments}

We thank MILC Collaboration for sharing the lattices used to perform this study. The LQCD calculations were performed using the Chroma software suite~\cite{Edwards:2004sx}. 
This research used resources of the National Energy Research Scientific Computing Center, a DOE Office of Science User Facility supported by the Office of Science of the U.S. Department of Energy under Contract No. DE-AC02-05CH11231 through ERCAP;
facilities of the USQCD Collaboration are funded by the Office of Science of the U.S. Department of Energy,
and supported in part by Michigan State University through computational resources provided by the Institute for Cyber-Enabled Research (iCER). 
The work of  ZF and HL is partially supported by the US National Science Foundation under grant PHY 1653405 ``CAREER: Constraining Parton Distribution Functions for New-Physics Searches'' and by the  Research  Corporation  for  Science  Advancement through the Cottrell Scholar Award.  The work of WG is supported by MSU University Distinguished Fellowship.
The work of HL is partially supported by the US National Science Foundation under grant PHY 2209424.


\end{document}